\documentclass[
aps,
prl,
amsmath,amssymb,
reprint,
superscriptaddress,
amsart,
]{revtex4-2}

\usepackage{graphicx}
\usepackage[utf8]{inputenc}
\graphicspath{ {images/} }
\usepackage{units}
\usepackage{xcolor}
\usepackage{tabularx}
\usepackage[percent]{overpic}
\usepackage{bm}

\usepackage{multirow}
\usepackage{verbatim}

\begin{document}

\title{
Comprehensive Acoustic Characterization of a Broadband Lateral MEMS Ultrasonic Transducer by Near-Field Acoustic Holography and Coupled Multiphysics Simulation
}

\author{Anton Melnikov}
\email{anton.melnikov@tum.de}
\affiliation{
Fraunhofer Institute for Photonic Microsystems IPMS, Dresden, Germany}

\author{Adrian Alexander}
\affiliation{
Fraunhofer Institute for Photonic Microsystems IPMS, Dresden, Germany}

\author{Tenia Meisel}
\affiliation{
Fraunhofer Institute for Photonic Microsystems IPMS, Dresden, Germany}

\author{Jorge M. Monsalve}
\affiliation{
Fraunhofer Institute for Photonic Microsystems IPMS, Dresden, Germany}




\begin{abstract}

Broadband airborne ultrasonic transducers are attractive for applications such as ranging, echo-location, and acoustic material characterization, where wide bandwidths enable advanced excitation and signal-processing techniques. Recently, a lateral capacitive micromachined ultrasonic transducer (L-CMUT) employing electrostatically actuated in-plane microbeams was introduced as a broadband MEMS transducer concept. However, a comprehensive investigation of its acoustic radiation behavior and quantitative validation of corresponding multiphysics models have not yet been reported.

In this work, we present a combined experimental and numerical characterization of a previously reported L-CMUT device operating between \unit[10]{kHz} and \unit[60]{kHz}. A coupled electrostatic--structural--thermoviscous finite-element model is combined with an acoustic boundary-element formulation to investigate the complete transduction chain from electrostatic actuation to acoustic radiation. The numerical predictions are validated experimentally using spherical near-field acoustic holography, enabling reconstruction of three-dimensional directivity patterns, radiated acoustic power, and spherical harmonic representations of the radiated sound field.

The results show that the broadband behavior originates from strong thermoviscous damping within the narrow beam--stator cavities, resulting in strongly overdamped operation despite a mechanical eigenfrequency near \unit[50]{kHz}. Good agreement between simulations and measurements is obtained for radiation patterns, radiated sound power, and spherical harmonic content. Furthermore, horn loading primarily redistributes acoustic energy and modifies directivity while only weakly affecting the total radiated acoustic power.

\end{abstract}

\maketitle

\section{Introduction}

Air-coupled micromachined ultrasonic transducers (MUTs) have become an enabling technology for a broad range of applications, including medical ultrasound \cite{Gerardo.2018,KhuriYakub.2011,Brenner.2019}, gesture recognition \cite{Wang.2019b,Gijsenbergh.2019}, object detection and ranging \cite{Yaralioglu.2003,Yu.2016,Przybyla.2015,Unger.2019}, and ultrasonic imaging \cite{Park.2013,KhuriYakub.2011}. Among the most impressive examples of airborne ultrasonic sensing found in nature are echolocating bats, which exploit broadband ultrasonic emissions to navigate complex environments and identify targets with extraordinary accuracy. Inspired by such biological systems, broadband airborne ultrasound has attracted growing interest for technical echo-location and ranging applications. In particular, chirped excitation signals and pulse-compression methods improve signal-to-noise ratio and measurement robustness while requiring transducers with sufficiently wide bandwidths \cite{claudon_advances_2002,bond_ultrasonics_2024,lacefield_ultrasound_2015,yamamoto_holographic_1984}. Broadband ultrasonic transducers have furthermore demonstrated advantages in applications such as simultaneous determination of thickness and speed of sound and non-contact identification of material properties \cite{bente_broadband_2023,otsuka_non-contact_2022}.

Despite this demand, the realization of broadband airborne ultrasonic transducers remains challenging. Conventional piezoelectric transducers are often narrowband and relatively bulky. Piezoelectric Micromachined Ultrasonic Transducers (PMUTs) provide compact dimensions and low power consumption but typically exhibit strongly resonant behavior \cite{Muralt.2005,Akasheh.2005,Przybyla2012}. Capacitive Micromachined Ultrasonic Transducers (CMUTs) enable CMOS-compatible fabrication and require very little drive current \cite{KhuriYakub.2011,Salim.2012}, making them attractive for integrated ultrasonic microsystems. However, many reported CMUTs operate at frequencies significantly above \unit[100]{kHz} \cite{Kupnik2011,ZhangH2020}. An early low-frequency CMUT operating near \unit[50]{kHz} was reported by Wygant \textit{et al.} \cite{Wygant50kHz}, but required bias voltages exceeding \unit[300]{V}.

Various approaches have been proposed to increase the bandwidth of airborne ultrasonic transducers. Kang \textit{et al.} experimentally demonstrated a wideband electrodynamic transducer operating between \unit[46.4]{kHz} and \unit[144.6]{kHz} \cite{kang_wideband_2019}. Ealo \textit{et al.} reported electromechanical film transducers driven below resonance and demonstrated broadband operation below \unit[150]{kHz} \cite{ealo_broadband_2008}. In the field of PMUTs, Sun \textit{et al.} realized a multi-frequency design with experimentally demonstrated bandwidths of 112\% and 38\% around \unit[615]{kHz} and \unit[1.63]{MHz}, respectively \cite{sun_investigation_2019}, while Wang \textit{et al.} proposed a multi-frequency PMUT combined with a Helmholtz resonator \cite{wang_multi-frequency_2023}. For CMUTs, bandwidth enhancement has been investigated through squeeze-film damping \cite{Apte2013,MaCMUT,Park2021}, rotational and rocking modes \cite{kuntzman_microfabrication_2015}, spiral geometries \cite{Adelegan_Spiral}, and arrays composed of multiple resonant cells \cite{zhang_fabrication_2020,wang_hybrid_2021,lu_theoretical_2022}. Although these approaches demonstrate broadband or multi-frequency behavior, many rely on multiple resonances, transducer arrays, or operation at comparatively high frequencies.

Besides frequency response, the spatial radiation characteristics of ultrasonic transducers are equally important for ranging and echo-location applications. Nevertheless, directivity is only sparsely discussed in the literature on broadband airborne ultrasonic transducers. Detailed directivity information has primarily been reported for the electrodynamic transducer of Kang \textit{et al.} \cite{kang_wideband_2019} and the electromechanical film transducers of Ealo \textit{et al.} \cite{ealo_broadband_2008}. For broadband MEMS ultrasonic transducers, comprehensive investigations of radiated sound power, full three-dimensional radiation patterns, and frequency-dependent beam shaping remain rare.

Recently, Monsalve \textit{et al.} introduced an alternative electrostatic transducer concept referred to as the lateral CMUT (L-CMUT) \cite{Monsalve.2022}. In contrast to conventional CMUTs, the device utilizes laterally oscillating electrostatic microbeams rather than membranes moving normal to the chip surface. The narrow cavity between beam and stator introduces strong squeeze-film damping \cite{Monsalve_2025_LRF}, resulting in a reduced quality factor and broadband behavior. Furthermore, the lateral arrangement enables the integration of a large number of microbeams within a compact footprint while maintaining moderate operating voltages. Initial demonstrations of fabrication, operation, and analytical modeling have already been reported \cite{Monsalve.2022}. Subsequently, Meisel \textit{et al.} demonstrated that acoustic horns can be attached to the device to improve directional radiation while largely preserving bandwidth \cite{Meisel2022}.

Despite these developments, a comprehensive experimental characterization of the acoustic radiation behavior of the L-CMUT has not yet been reported. In particular, the relationship between the internal thermoviscous dynamics of the transducer, the resulting radiated sound field, the radiated acoustic power, and the influence of acoustic beam-shaping elements remains insufficiently understood. Furthermore, quantitative validation of multiphysics acoustic models is challenging because the active microstructures are enclosed within the transducer and therefore inaccessible to direct optical characterization.

In this work, we present a comprehensive experimental and numerical characterization of a previously reported broadband L-CMUT device operating between \unit[10]{kHz} and \unit[60]{kHz}. A coupled electrostatic-structural-thermoviscous finite-element model is combined with a boundary-element radiation model to investigate the complete transduction chain from electrostatic actuation to acoustic radiation. The numerical predictions are validated experimentally using spherical near-field acoustic holography, enabling reconstruction of the full three-dimensional radiation field, radiated sound power, and spherical harmonic content. In addition, the acoustic behavior of a previously reported horn-loaded configuration is investigated. The results provide quantitative insight into the internal damping mechanisms, acoustic radiation characteristics, and beam-shaping behavior of broadband in-plane MEMS ultrasonic transducers.

\section{Results}

\subsection{Investigated L-CMUT device}

The investigated device is based on the lateral capacitive micromachined ultrasonic transducer (L-CMUT) platform previously reported by Monsalve \textit{et al.} \cite{Monsalve.2022}. In contrast to conventional CMUTs, the transducer employs electrostatically actuated microbeams oscillating parallel to the chip surface. Acoustic volume flow is generated inside narrow beam--stator cavities and radiated through dedicated outlet structures into the surrounding air.

\begin{figure*}[hbtp]
    \centering
    \textbf{a}\includegraphics[height = 7cm]{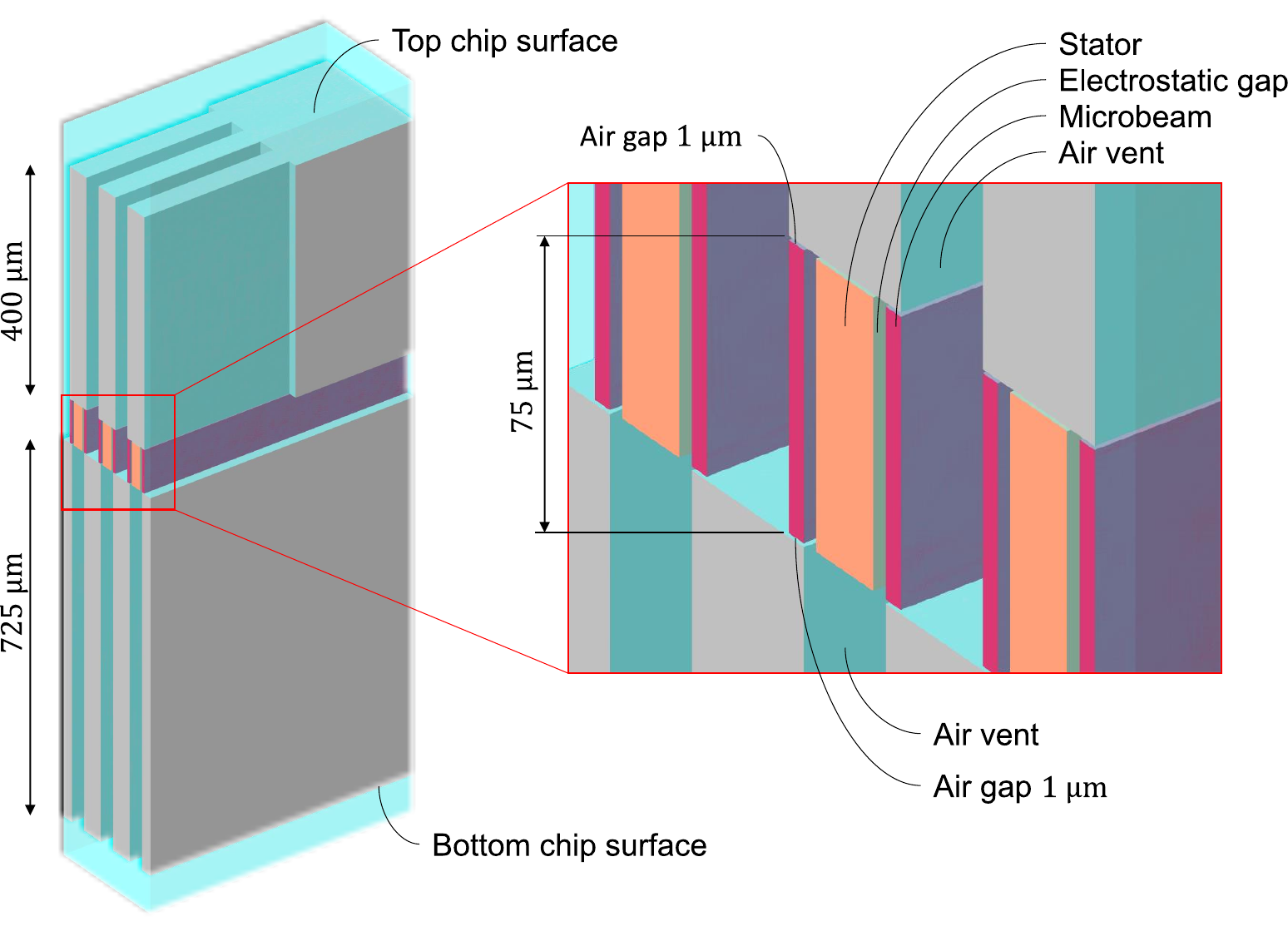}\\
    \textbf{b}\includegraphics[height = 5cm]{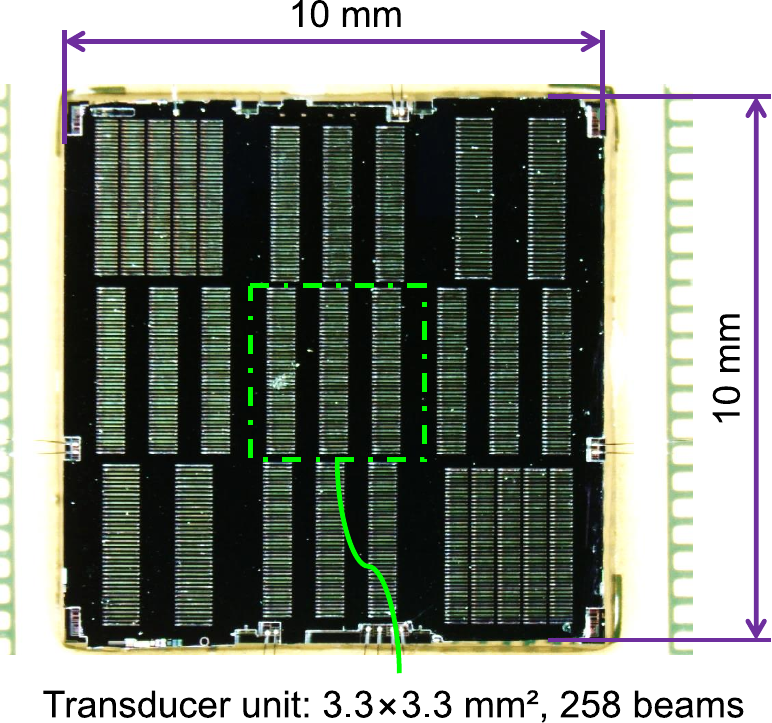}
    \textbf{c}\includegraphics[height = 5cm]{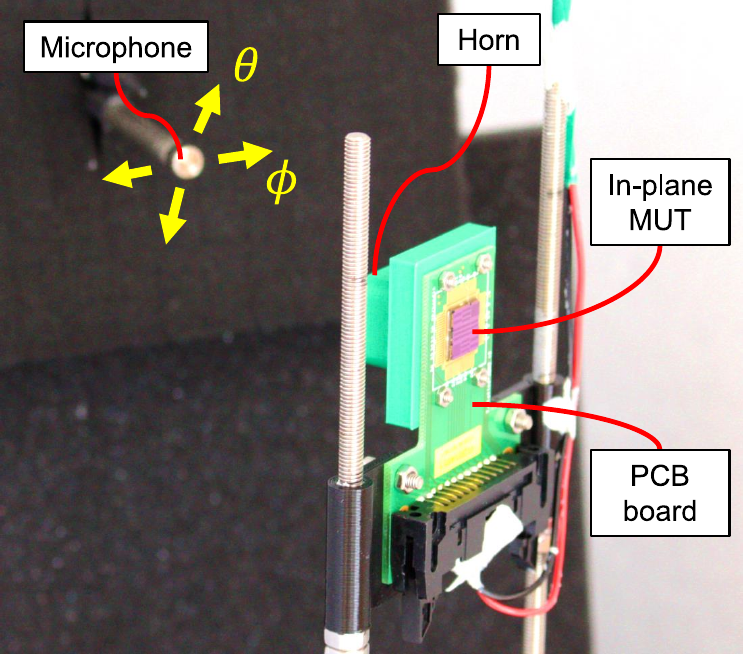}\\
    \textbf{d}\includegraphics[height = 7cm]{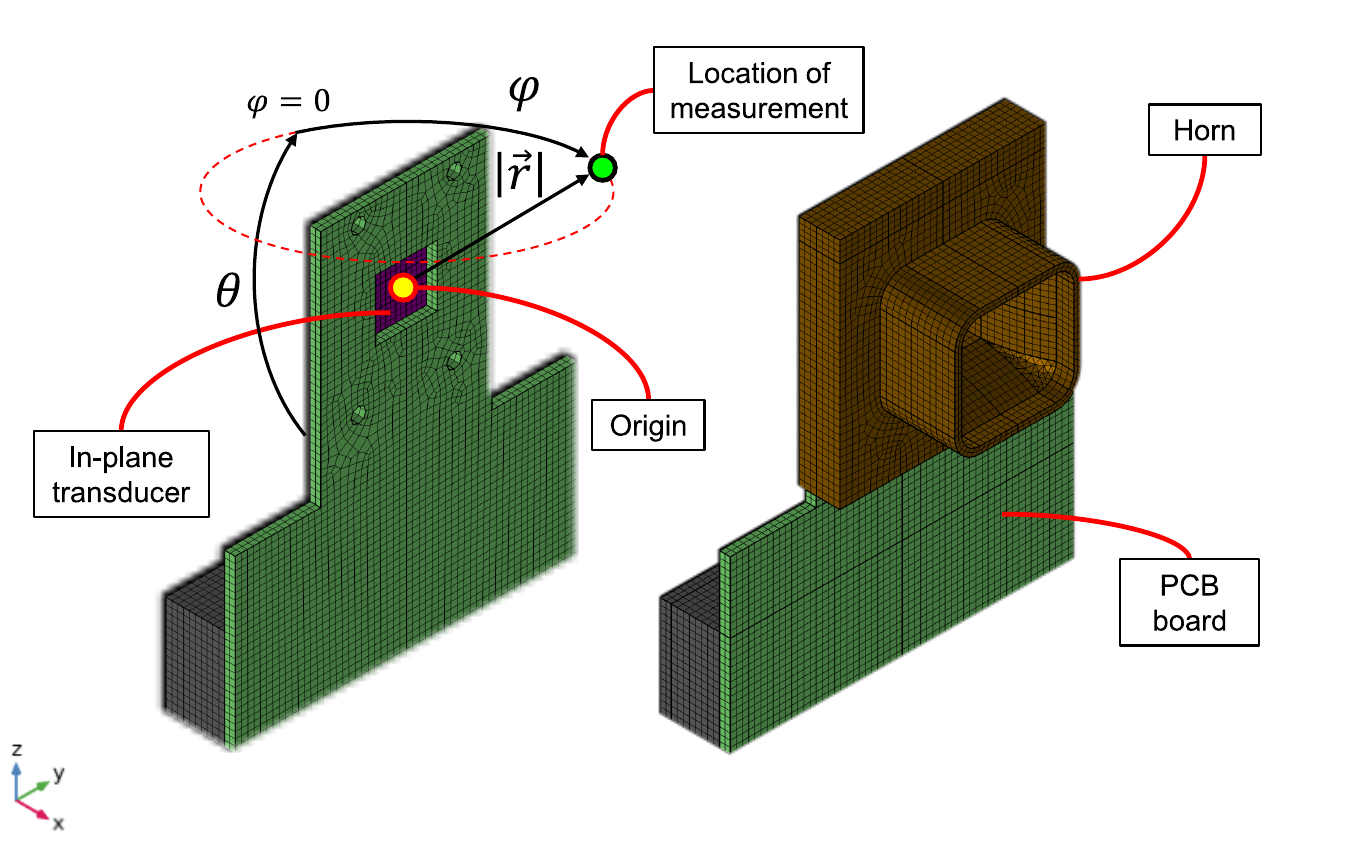}
    \caption{ Investigated L-CMUT platform and characterization approach. (a) Cross-sectional schematic of the investigated L-CMUT architecture. Electrostatic actuation of laterally oscillating microbeams generates acoustic volume flow within narrow beam--stator cavities, which is guided through vent structures and radiated from both sides of the chip. (b) Photograph of the fabricated chip containing nine transducer units arranged in a $3\times3$ configuration. The central transducer unit was used throughout this work. (c) Experimental setup employed for spherical near-field acoustic holography measurements of the unloaded and horn-loaded configurations. (d) Boundary-element models used for prediction of the radiated acoustic field, including the transducer, printed circuit board, socket, and horn where applicable. }
    \label{fig:first_principles}
\end{figure*}

Figure~\ref{fig:first_principles}a shows a cross-sectional schematic of the investigated L-CMUT architecture. Electrostatic actuation of laterally oscillating microbeams generates acoustic volume flow within narrow beam--stator cavities, which is subsequently guided through vent structures towards the top and bottom chip surfaces.
The transducer consists of a three-layer silicon structure comprising an active device layer enclosed by passive handle and lid layers. The active layer contains an array of electrostatic microbeam actuators, while the upper and lower layers guide the generated acoustic flow towards the outlet openings. A photograph of the fabricated transducer chip is shown in Fig.~\ref{fig:first_principles}b. The chip contains nine transducer units, each occupying an area of $\unit[3.3 \times 3.3]{mm^2}$. Throughout this work, only the central unit is considered in order to minimize the influence of edge effects and facilitate comparison between simulations and measurements.

Each transducer unit comprises 258 electrostatic microbeam actuators with a beam height of $h_b=\unit[75]{\mu m}$ and a beam length of $l_b=\unit[970]{\mu m}$. Owing to the lateral arrangement of the actuators, the acoustically active surface can be increased substantially relative to the chip footprint, enabling efficient acoustic excitation at moderate operating voltages. Detailed information regarding device design, fabrication, and initial characterization can be found in Ref.~\cite{Monsalve.2022} and is not repeated here.

To investigate the influence of acoustic beam shaping, a horn-loaded configuration is additionally considered. The horn design was previously proposed and optimized by Meisel \textit{et al.} \cite{Meisel2022}. A photograph of the measurement setup incorporating the horn-loaded transducer is shown in Fig.~\ref{fig:first_principles}c, while the corresponding numerical geometry employed for radiation simulations is shown in Fig.~\ref{fig:first_principles}d.

The objective of the present work is not to introduce a new transducer or horn design, but to establish a quantitative understanding of their acoustic behavior. To this end, coupled multiphysics simulations are combined with spherical near-field acoustic holography to investigate the complete transduction chain, from electrostatic actuation and internal thermoviscous flow generation to acoustic radiation and beam shaping.

\subsection{Bias-dependent electromechanical behavior}

The acoustic behavior of the L-CMUT is governed by the electromechanical operating point of the individual microbeam actuators. As previously shown by Monsalve \textit{et al.} \cite{Monsalve.2022}, electrostatic attraction between beam and stator results in a static pre-deflection that depends on the applied bias voltage. Since the stiffness, capacitance, and dynamic response of the actuator depend on this pre-deflected state, the operating point must be established before analyzing the acoustic performance of the device.\\

\begin{figure}[hbtp]
    \textbf{a}\\
    \includegraphics[width = \linewidth]{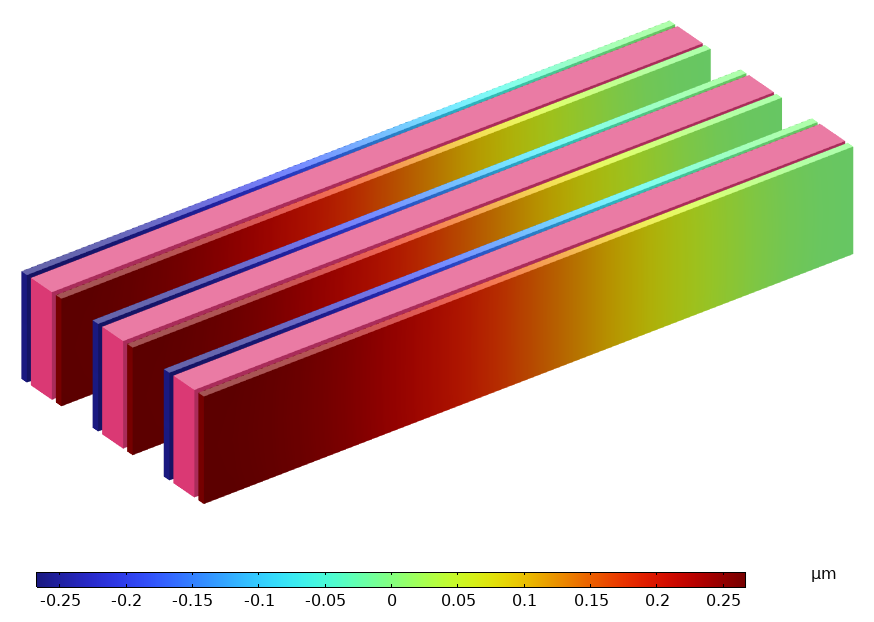}\\
    \textbf{b}\\
    \includegraphics[width = \linewidth]{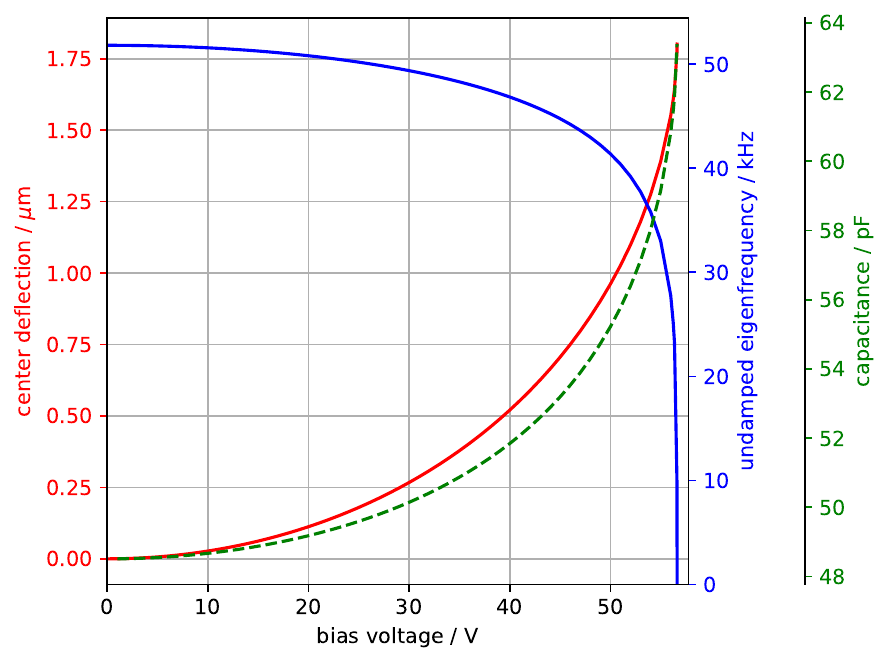}
    \caption{ \textbf{a} Equilibrium configuration of three neighboring L-CMUT cells at the operating bias voltage of \unit[30]{V}. The color scale represents the displacement in the actuation direction. Only one half of the geometry is shown for clarity. \textbf{b} Influence of the bias voltage on microbeam center deflection, first undamped eigenfrequency, and capacitance of the complete transducer unit. }
    \label{fig:bias_voltage}
\end{figure}

To determine the static operating conditions, a coupled electrostatic--structural finite-element model was solved for bias voltages between \unit[0]{V} and pull-in. Figure~\ref{fig:bias_voltage}a shows three neighboring L-CMUT cells in their deformed equilibrium configuration at the nominal operating bias of \unit[30]{V}. Only half of the geometry is shown. The color map illustrates the displacement in the actuation direction. At the operating bias voltage, the center deflection of the microbeam is calculated as \unit[267]{nm}.\\

The influence of bias voltage on the static center deflection is shown in Fig.~\ref{fig:bias_voltage}b. The deflection increases nonlinearly with increasing voltage and reaches \unit[1.8]{$\mu$m} at pull-in. The pull-in instability is predicted at \unit[56.6]{V}, providing a substantial safety margin with respect to the nominal operating voltage of \unit[30]{V}. Consequently, both the applied bias voltage and the additional signal voltage remain within the stable operating regime throughout this work.\\

Electrostatic softening further influences the dynamic properties of the actuator. Previous analytical models predict an undamped eigenfrequency of approximately \unit[54]{kHz} and a microbeam stiffness of approximately \unit[43]{N/m} \cite{Monsalve.2022,Melnikov.2021,Schenk.2022}. Figure~\ref{fig:bias_voltage}b shows the calculated first eigenfrequency of the microbeam as a function of bias voltage. At zero bias, the eigenfrequency obtained from the finite-element model is \unit[51.8]{kHz}.
Increasing electrostatic loading reduces the effective stiffness and causes a continuous decrease of the eigenfrequency, reaching \unit[49.4]{kHz} at the operating bias voltage of \unit[30]{V}. As expected for electrostatic actuators approaching pull-in, the eigenfrequency tends towards zero at the stability limit.\\

The capacitance of the complete transducer unit, comprising all 258 microbeam actuators, is also shown in Fig.~\ref{fig:bias_voltage}b. In contrast to previous analytical estimates \cite{Monsalve.2022}, the present model explicitly considers the recessed stator profile used to linearize the electrostatic force distribution. The resulting zero-bias capacitance is calculated as \unit[48.5]{pF}, corresponding to \unit[0.188]{pF} per actuator. At the nominal operating voltage of \unit[30]{V}, the capacitance increases to \unit[50.2]{pF} owing to electrostatic pre-deflection and reaches approximately \unit[63]{pF} at pull-in.\\

The electromechanical analysis establishes three important characteristics of the investigated device. First, stable operation is achieved with a substantial margin to pull-in. Second, electrostatic softening shifts the eigenfrequency only moderately within the considered voltage range. Third, the device operates in a significantly pre-deflected state, which directly influences the thermoviscous flow inside the beam--stator cavities and thus the resulting acoustic radiation. These effects are investigated in the following section.

\subsection{Thermoviscous dynamics inside the transducer}

To investigate the acoustic processes governing the broadband behavior of the L-CMUT, the previously described electromechanical model was extended by coupling thermoviscous acoustics to the structural mechanics domain. The resulting model accounts for viscous and thermal losses inside the narrow cavities and electrostatic gaps using the linearized Navier--Stokes equations. A fluid--structure interaction interface was used to couple the oscillating microbeams to the surrounding fluid. The acoustic domain additionally included small exterior volumes on the top and bottom surfaces of the chip to capture viscous and thermal boundary-layer effects at the outlet openings.\\

The frequency response was computed by linearizing the coupled system around the bias-dependent equilibrium state corresponding to an operating voltage of \unit[30]{V}. Harmonic simulations were performed between \unit[10]{kHz} and \unit[60]{kHz}. The resulting pressure field and acoustic velocity distributions provide direct insight into the mechanisms responsible for energy dissipation inside the transducer.\\

\begin{figure}[hbtp]
    \begin{tabular}{cc}
        \textbf{a} & \textbf{b}  \\
        \includegraphics[width = 0.48\linewidth]{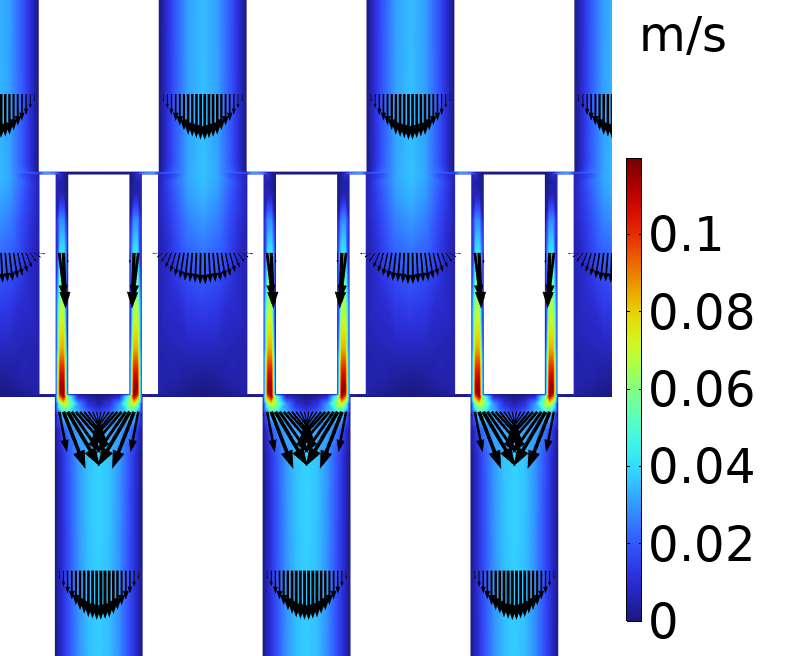} &
        \includegraphics[width = 0.48\linewidth]{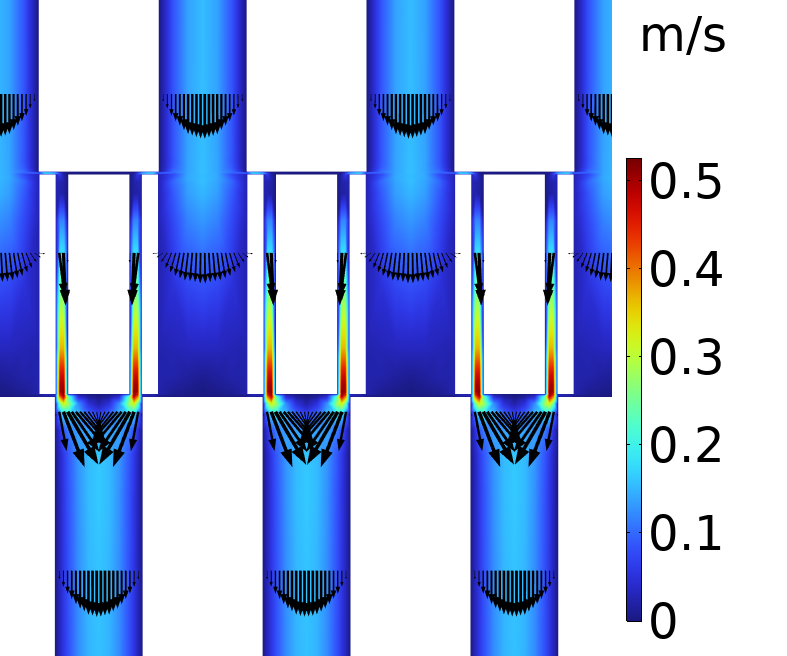}\\
         \textbf{c} & \textbf{d} \\
        \includegraphics[width = 0.48\linewidth]{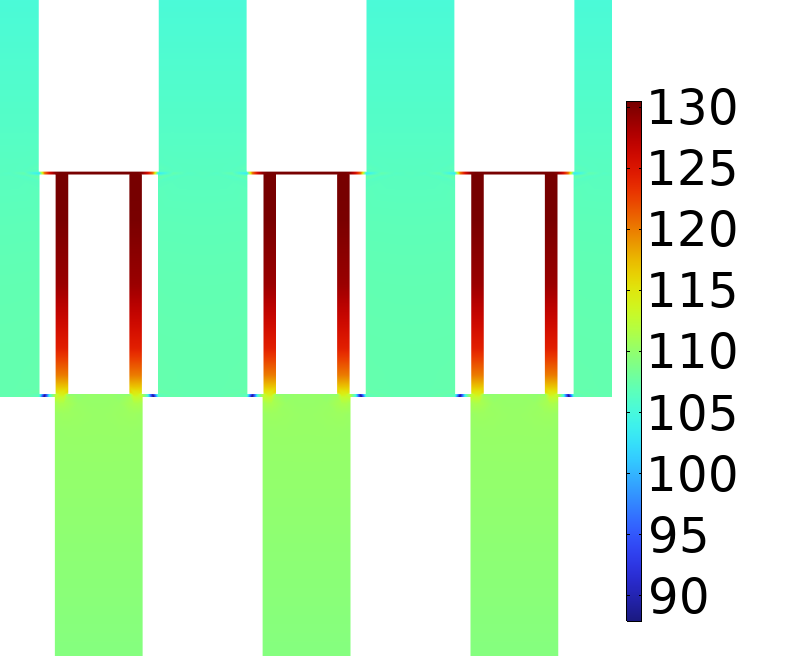} &
        \includegraphics[width = 0.48\linewidth]{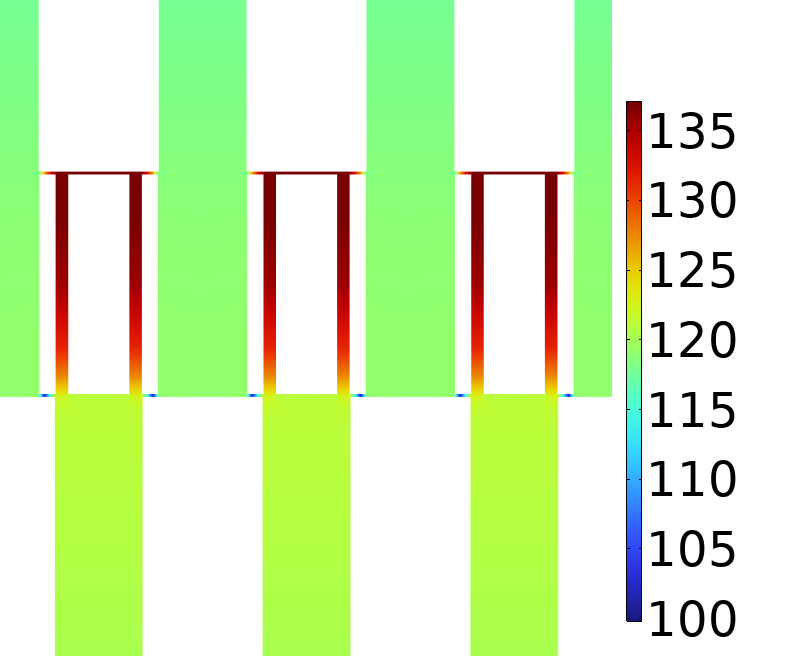}\\
    \end{tabular}
    \caption{ \textbf{a,b} Real part of the acoustic velocity field at \unit[10]{kHz} (\textbf{a}) and \unit[30]{kHz} (\textbf{b}). Arrows indicate the corresponding velocity vectors. The highest velocities occur within the narrow beam--stator cavities, where viscous dissipation is concentrated. \textbf{c,d} Sound pressure level distribution inside the transducer at \unit[10]{kHz} (\textbf{c}) and \unit[30]{kHz} (\textbf{d}). Elevated pressure levels remain largely confined to the internal fluidic structures and substantially exceed the radiated acoustic pressure. }
    \label{fig:microfluidics}
\end{figure}

Figures~\ref{fig:microfluidics}a and \ref{fig:microfluidics}b show the real part of the acoustic velocity field at \unit[10]{kHz} and \unit[30]{kHz}, respectively. The highest velocities occur inside the narrow beam--stator cavities, where the acoustic flow is constrained by the small gap dimensions. A pronounced velocity gradient can be observed across the gap height, indicating strong viscous losses. The flow profiles resemble Poiseuille-type motion, suggesting that the dominant damping mechanism originates from the confined fluid flow rather than structural dissipation within the silicon microbeams. At frequencies above approximately \unit[30]{kHz}, local flow velocities inside the gap approach \unit[0.5]{m/s}.\\

The pressure distributions corresponding to the same frequencies are shown in Figures~\ref{fig:microfluidics}c and \ref{fig:microfluidics}d. Owing to the restricted fluid motion inside the cavities, pressure levels substantially exceeding the radiated acoustic pressure are observed. At \unit[10]{kHz}, sound pressure levels inside the transducer reach approximately \unit[130]{dB}, while values around \unit[135]{dB} are predicted at \unit[30]{kHz}. These elevated pressure levels remain largely confined to the internal fluidic structures and therefore do not directly translate into far-field radiation.\\

The dimensions of the electrostatic gap place the flow regime close to the transition between continuum and slip-flow behavior. At atmospheric pressure, the Knudsen number exceeds $\mathrm{Kn}=0.01$ for characteristic dimensions below approximately \unit[6.5]{$\mu$m}. Consequently, slip-flow effects may influence the fluid dynamics inside the beam--stator cavities. Since no-slip boundary conditions are assumed, the predicted damping may be overestimated in the narrowest regions of the device.\\

\begin{figure}[hbtp]
        \includegraphics[width = 1\linewidth]{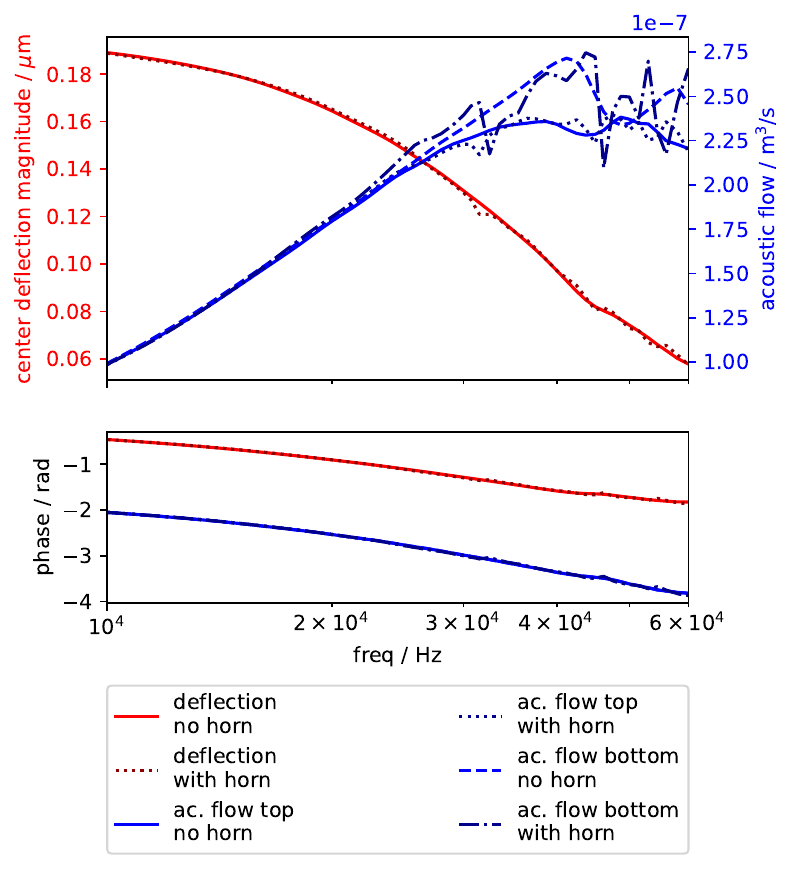} 
    \caption{ Simulated microbeam center displacement amplitude and acoustic volume flow emitted from the top and bottom sides of the transducer as a function of frequency for unloaded and horn-loaded configurations. Despite the first eigenfrequency being located near \unit[50]{kHz}, no pronounced resonance peak is observed in the displacement response. Instead, the transducer exhibits strongly overdamped behavior, while the acoustic volume flow increases with frequency and approaches a plateau above approximately \unit[40]{kHz}. Small differences between the top- and bottom-side volume flow become apparent at higher frequencies. The nearly identical responses obtained for the unloaded and horn-loaded configurations indicate that the horn has negligible influence on the internal electromechanical and thermoviscous dynamics of the transducer. }
    \label{fig:defl_ac_flow_freq}
\end{figure}

The influence of the internal fluid dynamics on the actuator response is illustrated in Fig.~\ref{fig:defl_ac_flow_freq}. The displacement magnitude at the center of the microbeam decreases monotonically from approximately \unit[190]{nm} at \unit[10]{kHz} to approximately \unit[60]{nm} at \unit[60]{kHz}. Remarkably, no pronounced resonance peak is observed, despite the eigenfrequency of the biased actuator being located near \unit[49.4]{kHz}. Instead, the response exhibits the behavior of a strongly overdamped system. Simultaneously, the acoustic volume flow increases with frequency and reaches a plateau above approximately \unit[40]{kHz}.
Above approximately \unit[30]{kHz}, small differences between the acoustic volume flow emitted from the top and bottom sides of the transducer become apparent, suggesting that compressibility effects within the chip become increasingly important at higher frequencies.\\

These observations indicate that the dynamic behavior of the L-CMUT is dominated by thermoviscous damping inside the beam--stator cavities. Rather than operating as a high-$Q$ resonator, the actuator behaves as a strongly damped acoustic flow source over a broad frequency range. This damping mechanism suppresses the resonance that would otherwise be expected from the mechanical eigenfrequency and therefore represents the physical origin of the broadband behavior of the investigated device.

\subsection{Coupling internal dynamics to acoustic radiation}

The previous section demonstrated that the dynamics of the L-CMUT are dominated by thermoviscous flow inside the beam--stator cavities. While these simulations provide detailed insight into the internal energy dissipation mechanisms, the resulting acoustic pressure and velocity fields cannot be directly observed experimentally. To enable comparison with measurements, the internally generated acoustic flow must be related to the radiated sound field outside the transducer.\\
\\
For this purpose, the thermoviscous finite-element model was coupled to an exterior acoustic model based on the boundary element method (BEM). The BEM geometry was derived from the CAD model of the investigated transducer assembly, including the MEMS chip, printed circuit board, socket, and horn where applicable. Geometric details whose dimensions were substantially smaller than the acoustic wavelength were omitted in order to reduce computational complexity. The resulting simulation domains are shown in Fig.~\ref{fig:first_principles}d. Owing to geometric symmetry, only half of the structure was considered during the calculations.\\
\\
The coupling between both models is based on the acoustic volume flow leaving the transducer through the outlet openings. The normal velocity obtained from the thermoviscous model was spatially averaged over the outlet boundary and transferred to the corresponding radiating surfaces of the BEM model. In turn, the acoustic pressure generated by the exterior domain was fed back to the thermoviscous model as a boundary condition. The coupling procedure enables a bidirectional interaction between the internal thermoviscous dynamics and the exterior acoustic field. Acoustic volume flow generated inside the transducer excites the surrounding medium, while the resulting acoustic loading is consistently accounted for through the coupled FEM--BEM formulation.\\
\\
The underlying assumption of this approach is that the array of closely spaced outlet openings behaves as an acoustically compact source region. Under this assumption, the detailed velocity distribution at individual outlet openings has only a minor influence on the radiated field compared to the total emitted volume flow. Nevertheless, local variations of radiation impedance across the chip surface are neglected. Consequently, the coupling should be regarded as an approximation whose validity is expected to improve when the characteristic outlet spacing remains small compared to the acoustic wavelength.\\
\\ 
The coupled simulation framework provides a direct link between the internal thermoviscous dynamics and the acoustic radiation into free space. In the following sections, the resulting radiation characteristics are compared with spherical near-field acoustic holography measurements in order to assess the accuracy of the predicted sound field.

\subsection{Experimental validation of the radiated sound field}

The coupled FEM--BEM model predicts the complete acoustic field radiated by the L-CMUT. To assess the validity of the model, the simulated radiation patterns were compared with spherical near-field acoustic holography measurements. The comparison was performed for both the bare transducer mounted on the PCB and the horn-loaded configuration described previously.\\

\begin{figure*}[hbtp]
    \centering
    \begin{tabular}{ll}
        \textbf{a} & \textbf{b}\\
        \includegraphics[width = 0.5\linewidth]{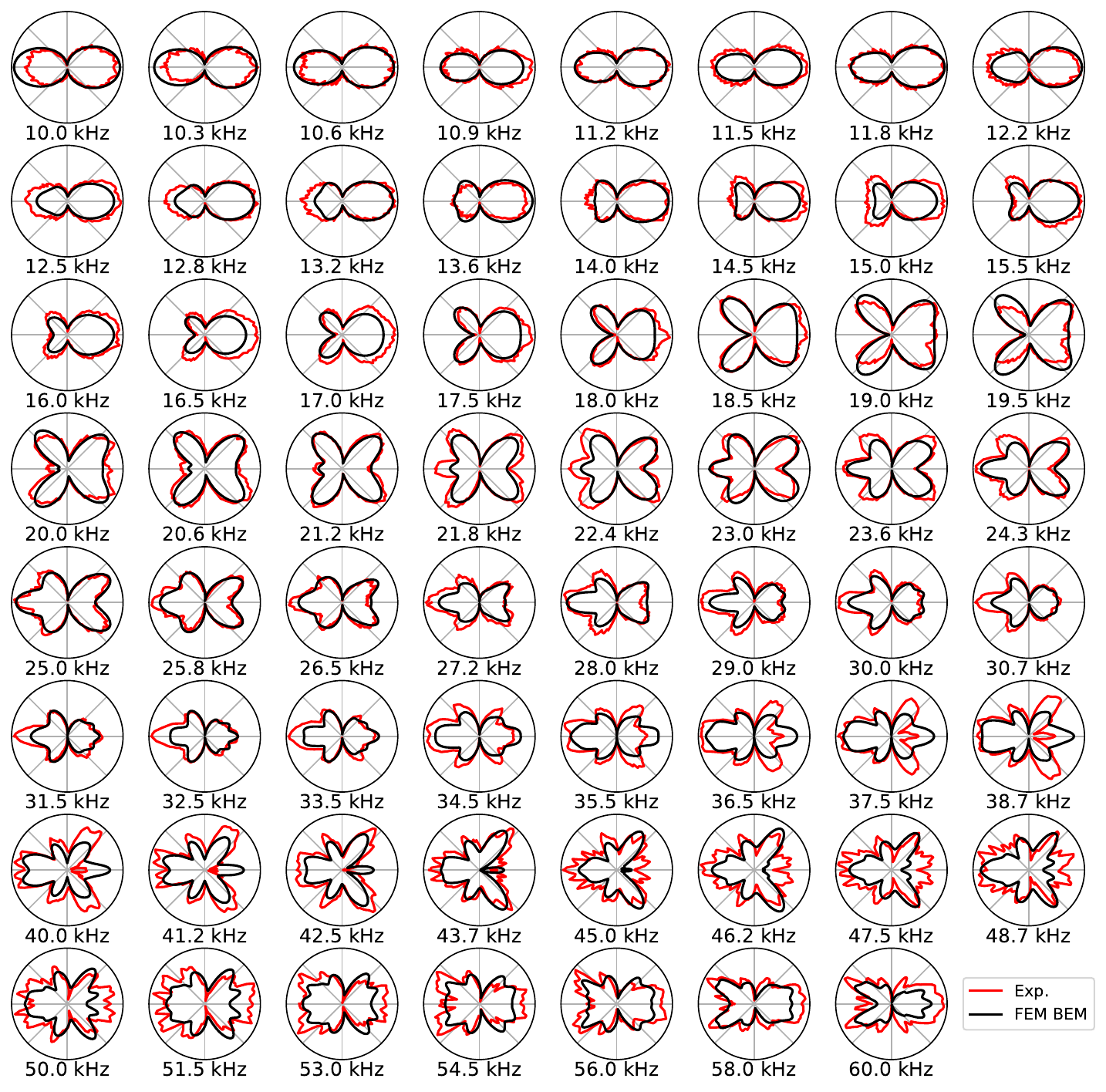} &
        \includegraphics[width = 0.5\linewidth]{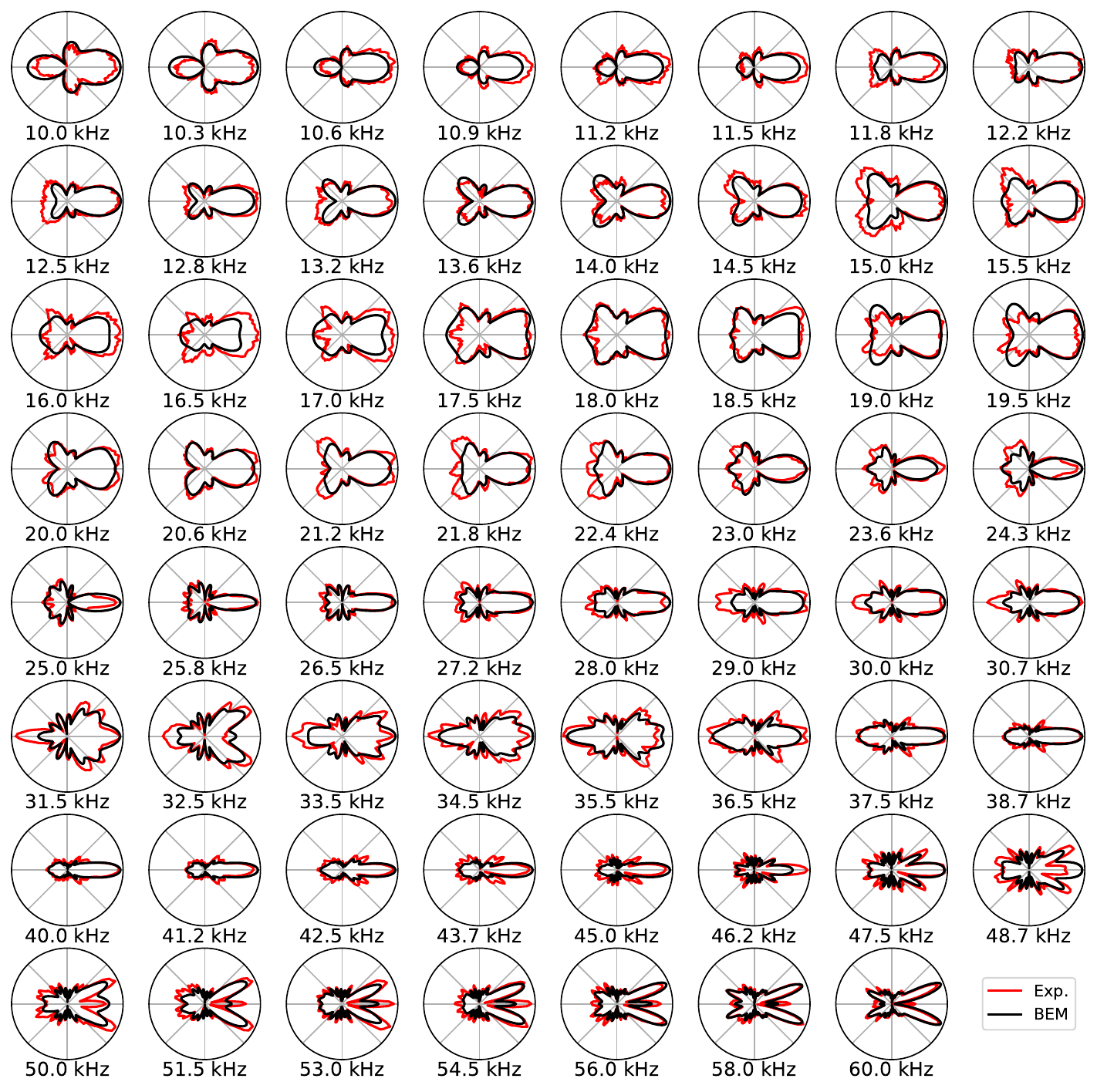} \\
        \textbf{c} & \textbf{d}\\
        \includegraphics[width = 0.5\linewidth]{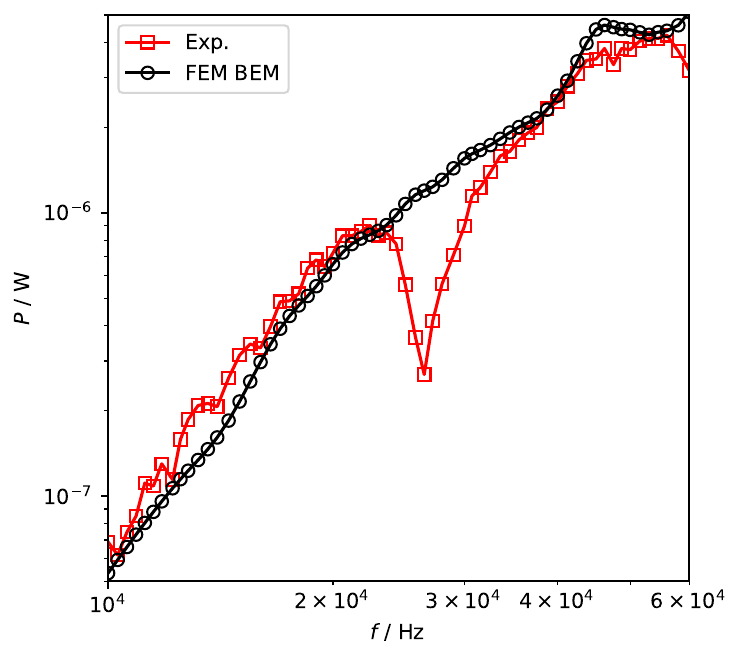} &
        \includegraphics[width = 0.5\linewidth]{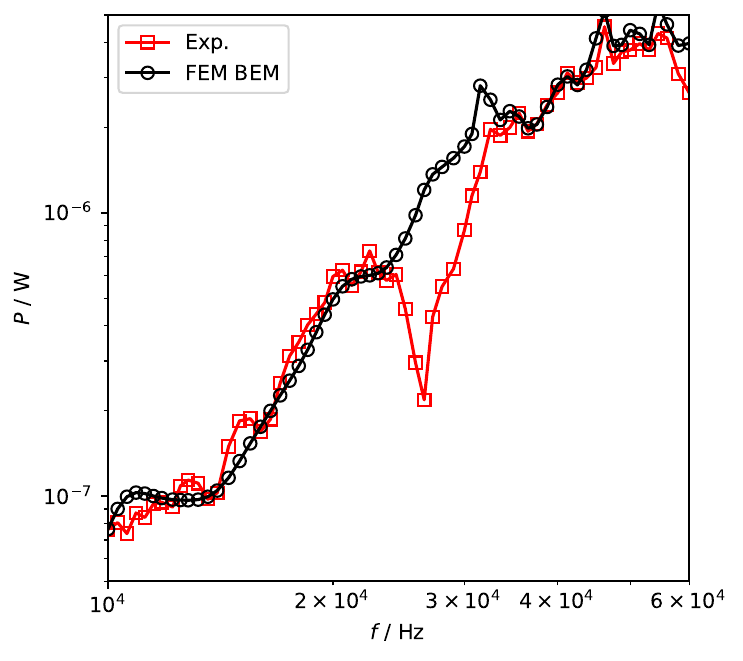} \\
    \end{tabular}
    \caption{ Comparison of measured and simulated radiation characteristics for unloaded and horn-loaded transducer configurations. \textbf{a} Frequency-dependent directivity patterns in the $xy$-plane for the unloaded transducer obtained from spherical near-field acoustic holography and FEM--BEM simulations. \textbf{b} Frequency-dependent directivity patterns in the $xy$-plane for the horn-loaded transducer obtained from spherical near-field acoustic holography and FEM--BEM simulations. \textbf{c} Radiated acoustic power of the unloaded transducer determined from spherical harmonic reconstruction and FEM--BEM simulations. \textbf{d} Radiated acoustic power of the horn-loaded transducer determined from spherical harmonic reconstruction and FEM--BEM simulations. }
    \label{fig:empty_results}
\end{figure*}

Figure~\ref{fig:empty_results}a shows the measured and simulated directivity patterns in the $xy$-plane for the transducer without horn. Across the investigated frequency range, the numerical model reproduces the main features of the measured acoustic field. At lower frequencies, the radiation is characterized by two dominant lobes located on opposite sides of the chip. With increasing frequency, the directivity pattern becomes progressively more complex and additional side lobes emerge. The measured and simulated lobe positions show good agreement throughout most of the investigated frequency range.\\

Small deviations between simulation and experiment are observed, particularly between approximately \unit[35]{kHz} and \unit[40]{kHz}. Since the corresponding wavelengths are comparable to several geometrical details of the measurement setup, these discrepancies are likely related to simplifications introduced in the numerical model, including omitted screws, connectors, wiring, and manufacturing tolerances of the assembled system.\\

The corresponding results for the horn-loaded configuration are shown in Fig.~\ref{fig:empty_results}b. Compared to the unloaded transducer, the radiation pattern is significantly modified and a pronounced main lobe develops along the horn axis. The simulations accurately reproduce both the orientation and the frequency dependence of the dominant lobe. Furthermore, the measured and simulated side-lobe structures show good qualitative agreement over a broad frequency range.\\

The agreement between simulations and measurements demonstrates that the coupled thermoviscous FEM--BEM model captures the dominant radiation mechanisms of the investigated device. Since the internal microbeam motion and cavity acoustics are not experimentally accessible, the radiated sound field provides an indirect validation of the predicted internal dynamics. In particular, the accurate reproduction of the frequency-dependent directivity patterns indicates that both the generated acoustic volume flow and its radiation into the surrounding medium are described consistently by the model.

\subsection{Radiated acoustic power}

While the agreement of the directivity patterns provides confidence in the predicted radiation characteristics, directivity alone does not guarantee that the radiated acoustic energy is reproduced correctly. A more stringent validation is obtained by comparing the radiated sound power, since it represents the total acoustic energy emitted by the transducer and is largely independent of local interference effects.\\
\\
Figures~\ref{fig:empty_results}c and \ref{fig:empty_results}d compare the radiated sound power obtained from spherical near-field acoustic holography with the corresponding FEM--BEM predictions for the unloaded and horn-loaded configurations, respectively. In both cases, the measured and simulated results show a similar overall frequency dependence. Starting from \unit[10]{kHz}, the radiated power gradually increases with frequency and reaches its maximum values in the upper part of the investigated frequency range.\\

For the unloaded transducer, the numerical model reproduces the measured power levels throughout most of the investigated frequency range. The largest discrepancy is observed around \unit[26.7]{kHz}, where the experimental data exhibits a pronounced notch that is not fully reproduced by the simulations. Since this feature appears consistently in the measurements, it is likely associated with physical effects that are not captured by the current model, such as geometric asymmetries, assembly tolerances, or additional acoustic interactions within the measurement setup.\\

A similar behavior is observed for the horn-loaded configuration. The overall agreement between simulation and experiment remains good, while the notch near \unit[26.7]{kHz} is again primarily visible in the experimental results. Outside this frequency region, the numerical predictions successfully reproduce both the magnitude and frequency dependence of the radiated power.\\

Compared to local pressure measurements, radiated sound power constitutes a particularly demanding validation metric because it depends on the acoustic field over the entire measurement surface. The observed agreement therefore indicates that the coupled thermoviscous FEM--BEM model not only reproduces the directivity of the radiated sound field but also predicts the total emitted acoustic energy with good accuracy.\\

Having established agreement in both directivity and radiated power, the measured sound fields can subsequently be analyzed using a spherical harmonic representation. This provides additional insight into the frequency-dependent evolution of the radiation characteristics and the influence of the horn on the emitted acoustic field.

\subsection{Spherical harmonic representation of the radiated field}

The previous sections demonstrated that the coupled FEM--BEM model accurately reproduces both the directivity patterns and the radiated acoustic power of the investigated platform. To obtain additional insight into the underlying radiation behavior, the measured and simulated sound fields were subsequently analyzed using a spherical harmonic representation.\\

\begin{figure*}[hbtp]
    \centering
    \textbf{a}\includegraphics[height = 5cm]{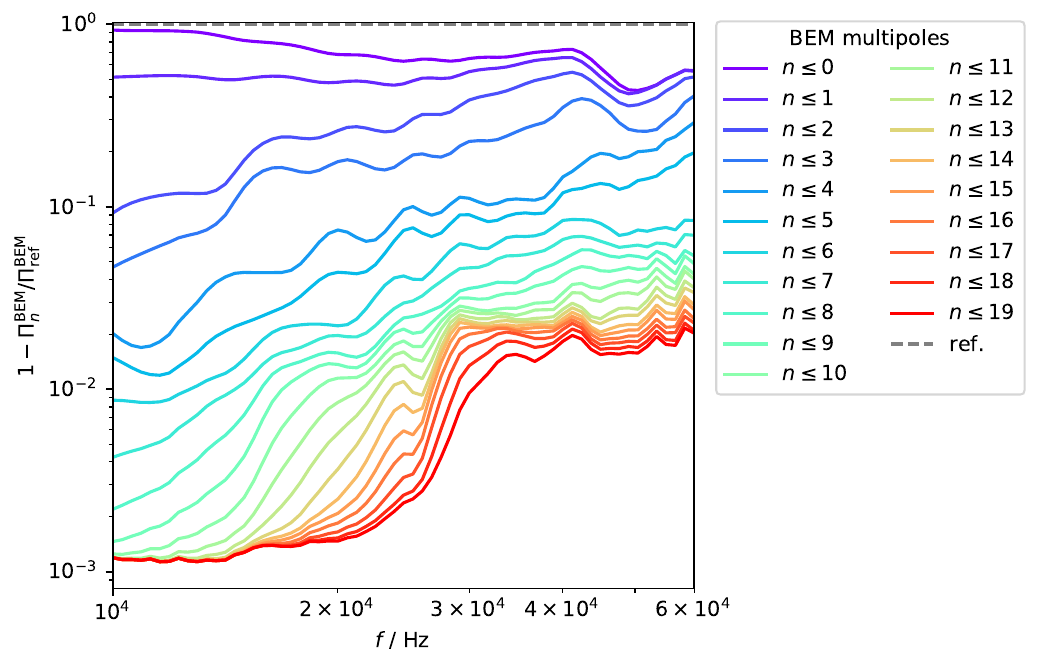}
    \textbf{b}\includegraphics[height = 5cm]{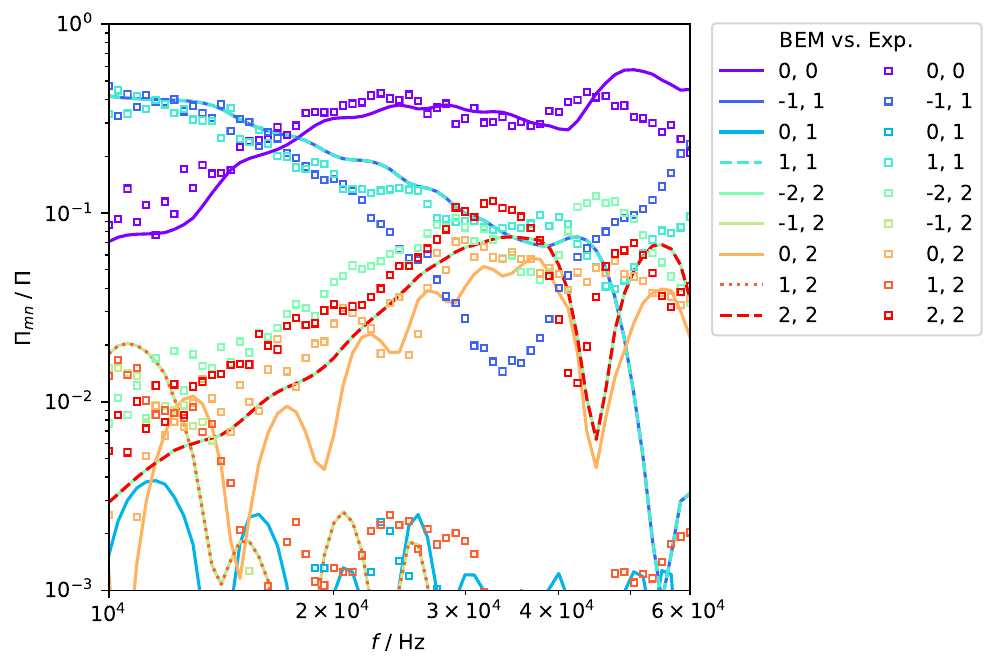}\\
    \textbf{c}\includegraphics[height = 5cm]{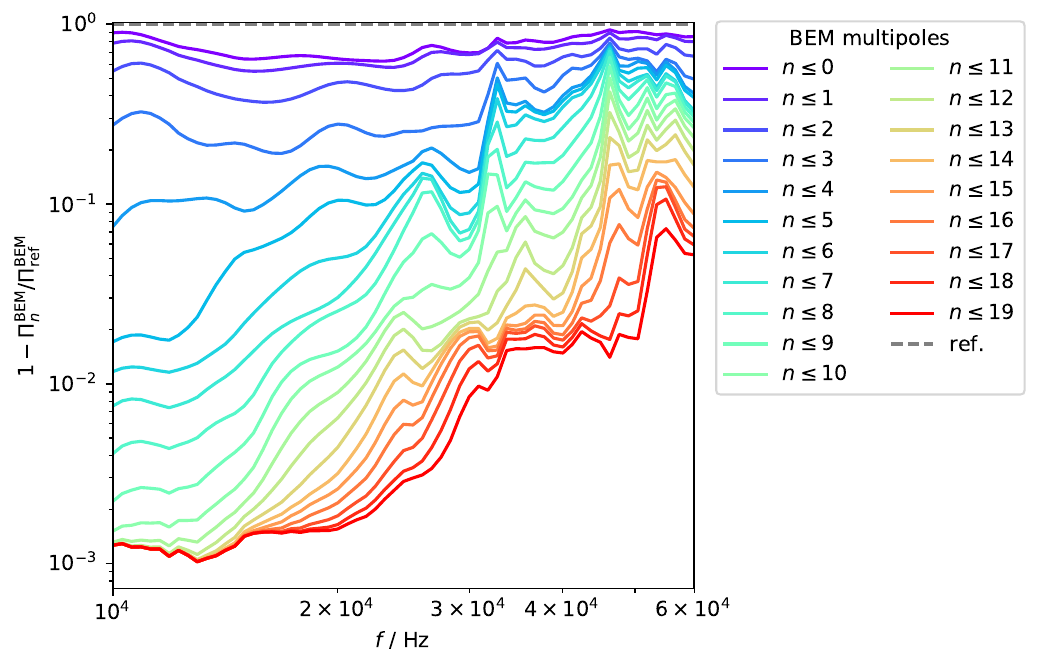}
    \textbf{d}\includegraphics[height = 5cm]{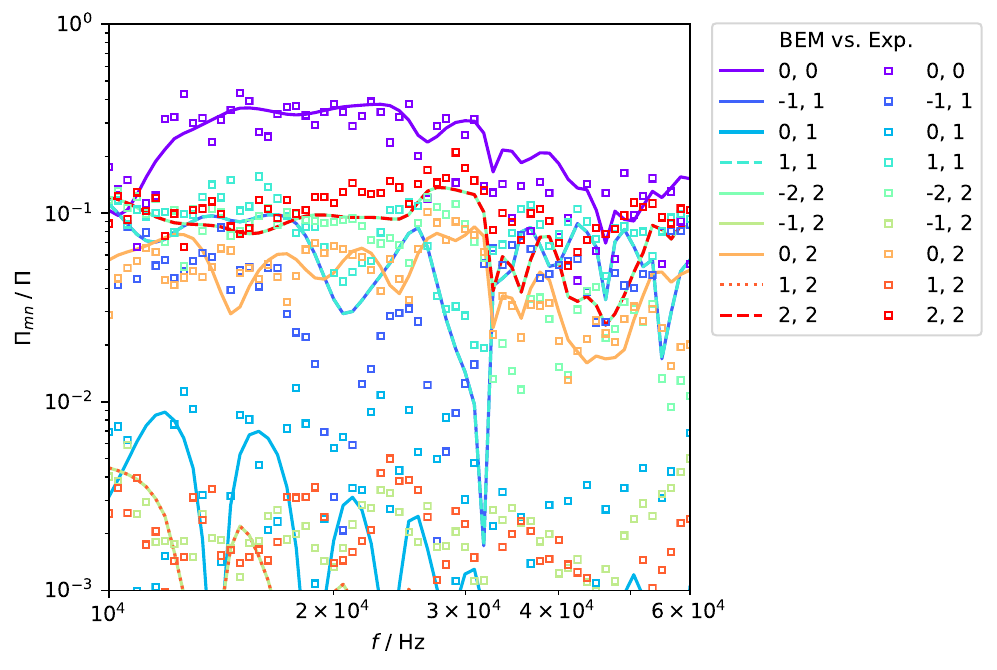}
    \caption{ Spherical harmonic representation of the radiated sound field for unloaded and horn-loaded transducer configurations. \textbf{a} Relative error of the reconstructed radiated acoustic power as a function of frequency for different spherical harmonic truncation orders in the unloaded configuration. The error is evaluated relative to the FEM--BEM reference solution. \textbf{b} Normalized power associated with the dominant spherical harmonic coefficients obtained from spherical near-field acoustic holography measurements and FEM--BEM simulations for the unloaded transducer. \textbf{c} Relative error of the reconstructed radiated acoustic power as a function of frequency for different spherical harmonic truncation orders in the horn-loaded configuration. The error is evaluated relative to the FEM--BEM reference solution. \textbf{d} Normalized power associated with the dominant spherical harmonic coefficients obtained from spherical near-field acoustic holography measurements and FEM--BEM simulations for the horn-loaded transducer. }
    \label{fig:coef_exp_vs_BEM_log}
\end{figure*}

Prior to evaluating the spherical harmonic coefficients, the convergence of the reconstructed sound power was investigated. The relative error associated with truncating the spherical harmonic expansion at different orders is shown in Fig.~\ref{fig:coef_exp_vs_BEM_log}a for the unloaded transducer and in Fig.~\ref{fig:coef_exp_vs_BEM_log}c for the horn-loaded configuration. As expected, the error increases with frequency due to the increasing spatial complexity of the radiated field. Using harmonics up to order $n=19$, corresponding to 400 complex coefficients, the relative power error remains below approximately \unit[2.2]{\%} over most of the investigated frequency range. This truncation order was therefore adopted throughout the subsequent analysis.\\

Figure~\ref{fig:coef_exp_vs_BEM_log}b shows the normalized power associated with the dominant spherical harmonic coefficients for the unloaded transducer. At lower frequencies, the radiated field is largely represented by the coefficients corresponding to $n=1$, indicating a radiation pattern dominated by two opposing main lobes. With increasing frequency, energy is progressively redistributed towards higher-order coefficients, reflecting the increasing spatial complexity of the radiated sound field observed in the directivity measurements.\\

The measured and simulated coefficient distributions show good qualitative agreement throughout the investigated frequency range. In particular, both datasets predict similar frequency regions where individual coefficients become dominant or diminish. Furthermore, the symmetry of the transducer geometry is reflected in the approximately equal power associated with coefficient pairs corresponding to positive and negative azimuthal orders. Small deviations from these symmetries are observed in the measurements, particularly at higher frequencies, which may be attributed to manufacturing tolerances, assembly imperfections, or positioning uncertainties during the holographic reconstruction.\\

The corresponding convergence results are shown in Fig.~\ref{fig:coef_exp_vs_BEM_log}c, while the normalized spherical harmonic coefficients of the horn-loaded configuration are presented in Fig.~\ref{fig:coef_exp_vs_BEM_log}d. Compared to the unloaded transducer, the spherical harmonic power is distributed across a larger number of coefficients over much of the investigated frequency range. This behavior reflects the modified radiation pattern introduced by the horn and the associated redistribution of acoustic energy into the forward direction. As frequency increases, progressively higher-order coefficients contribute to the radiated field, consistent with the increasingly structured directivity patterns observed experimentally.\\

The spherical harmonic representation therefore provides a compact description of the frequency-dependent radiation behavior of both configurations. Moreover, the good agreement between measured and simulated coefficient distributions further supports the validity of the coupled FEM--BEM approach and its ability to capture the dominant radiation mechanisms of the investigated L-CMUT device.

\subsection{Acoustic energy redistribution by horn loading}

The previous sections demonstrated that horn loading significantly modifies the directivity of the radiated sound field. A remaining question is whether the observed increase in sound pressure along the horn axis originates from an increase in radiated acoustic power or from a redistribution of the existing acoustic energy.\\

\begin{figure}[hbtp]
        \includegraphics[width = 1\linewidth]{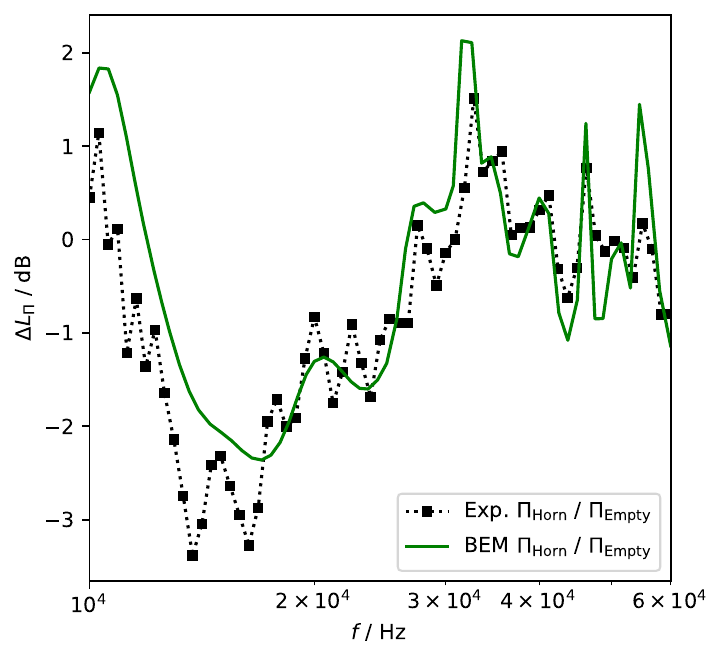} 
    \caption{ Change in radiated acoustic power level introduced by horn loading. Experimental results obtained from spherical near-field acoustic holography are compared with FEM--BEM simulations. Positive values indicate an increase in radiated acoustic power relative to the unloaded configuration, whereas negative values indicate a reduction. The overall power difference remains small across most of the investigated frequency range, indicating that the horn primarily redistributes the emitted acoustic energy rather than increasing the generated acoustic power. }
    \label{fig:power_empty_vs_horn}
\end{figure}

Figure~\ref{fig:power_empty_vs_horn} compares the change in radiated sound power level introduced by the horn. Experimental results obtained from spherical near-field acoustic holography are shown together with the corresponding FEM--BEM predictions. Both datasets exhibit a similar frequency dependence. Below approximately \unit[30]{kHz}, the horn reduces the radiated power by up to \unit[3]{dB}, while only minor power differences are observed throughout the upper part of the investigated frequency range.\\

The close agreement between simulation and experiment indicates that the coupled thermoviscous FEM--BEM model correctly captures the influence of the horn on the global radiation characteristics of the system. In particular, both approaches predict that the horn only weakly affects the total radiated acoustic power despite substantially modifying the radiation pattern.\\

These results indicate that the primary function of the horn is not to increase acoustic power generation by the transducer. Instead, the horn redistributes the emitted acoustic energy and concentrates it into a narrower angular region. Consequently, significant sound pressure gains can be achieved in the desired radiation direction without a corresponding increase in total radiated acoustic power.

\section{Discussion}

A distinctive characteristic of the investigated L-CMUT is that its useful operating bandwidth extends over a frequency range that includes the mechanical eigenfrequency without exhibiting the pronounced resonance peak typically associated with resonant ultrasonic transducers. This behavior contrasts with many PMUT and CMUT designs, where the achievable bandwidth is fundamentally linked to the quality factor of a dominant mechanical mode. In the present device, the acoustic response is instead governed by the interaction between the electrostatic actuator and the confined fluid domain.\\
The simulations indicate that the narrow beam--stator cavities introduce substantial thermoviscous dissipation, particularly inside the electrostatic gaps where the largest velocity gradients occur. Consequently, acoustic energy is continuously dissipated within the fluidic domain, suppressing resonant amplification of the structural motion. While such losses would generally be regarded as detrimental in resonant ultrasonic transducers, they become advantageous in the present context because they provide a broadband response without requiring multiple resonances, additional damping structures, or complex excitation schemes.\\
From a system perspective, this behavior is particularly attractive for applications relying on broadband excitation signals such as chirps or coded waveforms. Instead of maximizing the response at a single frequency, the L-CMUT provides a comparatively uniform acoustic output over a broad frequency range, enabling efficient utilization of advanced signal-processing techniques and potentially improving ranging robustness and target discrimination.

A second important outcome of this work is the successful validation of the coupled multiphysics model through measurements of the radiated acoustic field. Unlike many conventional ultrasonic transducers, the active acoustic elements of the L-CMUT are enclosed within the silicon structure and are therefore inaccessible to direct optical characterization techniques such as laser Doppler vibrometry. Consequently, validation based solely on structural measurements is not feasible.\\
Instead, the present study adopts an acoustics-driven validation strategy in which the complete radiated sound field is reconstructed using spherical near-field acoustic holography and compared with the predictions of a coupled electrostatic--structural--thermoviscous FEM--BEM model. Agreement was observed not only for local directivity patterns but also for radiated sound power and the spherical harmonic representation of the measured field. Since these quantities are influenced by the entire transduction chain, from electrostatic actuation to acoustic radiation, their agreement provides considerably stronger validation than a comparison of isolated response quantities.\\
Naturally, agreement of the radiated field does not prove that every detail of the internal flow field is predicted perfectly. Nevertheless, the observed consistency between measured and simulated directivity patterns, radiated sound power, and spherical harmonic content indicates that the dominant physical mechanisms governing acoustic generation and radiation are captured correctly. The presented methodology therefore provides a practical framework for the validation of enclosed MEMS acoustic transducers, where direct access to the active structures is limited or entirely unavailable.\\

The horn-loaded configuration further illustrates the distinction between sound pressure and radiated acoustic power. While the measured directivity patterns demonstrate a substantial increase in pressure along the horn axis, the corresponding sound-power analysis shows only minor changes in the total acoustic energy radiated by the transducer. The observed pressure gain therefore cannot be attributed to enhanced acoustic power generation. Instead, it originates primarily from a redistribution of the emitted acoustic energy into a narrower angular region.\\
From an acoustical perspective, the horn acts as a radiation-transforming element that modifies the spatial distribution of the radiated field rather than the energy generated at the source. Consequently, the horn increases the acoustic intensity in selected directions while reducing it elsewhere. This behavior explains why pressure gains of several tens of decibels can be achieved without a comparable increase in radiated power. The distinction is particularly relevant for ultrasonic ranging and echo-location applications, where detection performance is often governed by signal level in a limited angular region rather than by total acoustic output.\\
An additional consequence is that the acoustic performance of the complete system cannot be assessed solely from local pressure measurements. Depending on the application, a strong increase in on-axis pressure may create the impression of a more efficient transducer despite little change in total radiated energy. The combination of directivity measurements and radiated sound-power analysis employed in this work therefore provides a more complete characterization of the acoustic behavior of horn-loaded ultrasonic MEMS systems.\\

Finally, several limitations of the present study should be noted. The thermoviscous model assumes continuum flow and no-slip boundary conditions throughout the fluid domain. However, the characteristic dimensions of the electrostatic gaps approach the slip-flow regime, where rarefaction effects may become relevant. Consequently, the fluidic damping inside the narrowest regions of the device may be overestimated. Furthermore, the FEM--BEM coupling assumes an acoustically compact source region and neglects spatial variations in radiation impedance across individual outlet openings. While these approximations appear adequate for the investigated frequency range, their influence may become more pronounced for future device generations operating at higher frequencies or with smaller characteristic dimensions.\\
Additional discrepancies between simulation and experiment are observed around \unit[26.7]{kHz}, where a pronounced notch appears in the measured radiated sound power but is not reproduced by the numerical model. Since the feature is observed at approximately the same frequency for both the unloaded and horn-loaded configurations, it is unlikely to originate from the horn itself or from the external acoustic radiation. Likewise, no corresponding feature is observed in the coupled thermoviscous FEM--BEM simulations. One possible explanation is the presence of spatially non-uniform actuator excitation across the transducer unit, resulting from distributed electrical or structural effects that are not represented by the unit-cell model. Such mechanisms could introduce frequency-dependent variations in amplitude or phase between different regions of the transducer and thereby reduce the coherent acoustic output. Further investigations are required to assess the origin of this behavior.\\
Despite these limitations, the combined use of thermoviscous FEM, acoustic BEM, and spherical near-field acoustic holography provides a consistent description of the transduction chain from electrostatic actuation to acoustic radiation. Future work may extend the present framework by incorporating slip-flow boundary conditions, investigating alternative acoustic loading concepts, and applying the methodology to receiving as well as transmitting MEMS ultrasonic transducers.

\section{Conclusion} In this work, a comprehensive experimental and numerical characterization of a previously reported broadband lateral capacitive micromachined ultrasonic transducer (L-CMUT) platform was presented. A coupled electrostatic--structural--thermoviscous finite-element model was combined with an acoustic boundary-element formulation to investigate the complete transduction chain from electrostatic actuation to acoustic radiation.\\
The simulations demonstrated that the dynamic behavior of the device is dominated by thermoviscous losses within the narrow beam--stator cavities. As a consequence, no pronounced resonance peak is observed despite the mechanical eigenfrequency being located near \unit[50]{kHz}. Instead, the transducer operates as a strongly damped acoustic source over a broad frequency range extending from \unit[10]{kHz} to \unit[60]{kHz}.\\
The numerical predictions were validated experimentally using spherical near-field acoustic holography. Good agreement was obtained for the directivity patterns, radiated acoustic power, and spherical harmonic representation of the radiated sound field. The results demonstrate that full-field acoustic measurements provide an effective approach for validating multiphysics models of enclosed MEMS ultrasonic transducers, where direct access to the active structures is limited.\\
Furthermore, the influence of a previously reported horn-loaded configuration was investigated. While the horn significantly modified the directivity pattern and increased sound pressure along the preferred radiation direction, only minor changes in total radiated acoustic power were observed. The results therefore indicate that the primary function of the horn is the redistribution of acoustic energy rather than an increase in acoustic power generation.\\
The presented combination of multiphysics simulations and spherical acoustic holography provides a framework for the analysis and optimization of broadband airborne MEMS ultrasonic transducers and may be extended to future transmitter and receiver architectures employing enclosed acoustic microstructures.

\section{Materials and methods}

\subsection{Thermoviscous FEM model}
The internal electromechanical and acoustic behavior of the previously reported L-CMUT platform \cite{Monsalve.2022} was investigated using a three-dimensional finite-element model implemented in COMSOL Multiphysics. The model combines electrostatics, structural mechanics, thermoviscous acoustics, fluid--structure interaction, and moving mesh formulations in order to capture the complete interaction between the electrostatic actuator and the surrounding fluid. Owing to geometric symmetry, only half of a single actuator cell was modeled. This reduction decreases the computational effort by reducing the number of degrees of freedom while maintaining sufficient spatial resolution within the electrostatic gaps and microacoustic flow channels.\\
The stationary operating point was determined using a nonlinear electrostatic--structural analysis. Electrostatic forces and structural deformation were solved self-consistently while accounting for geometric nonlinearities through a moving mesh formulation. The resulting equilibrium solution provided the bias-dependent actuator deflection, capacitance, and effective stiffness.\\
Subsequently, the complete multiphysics system was linearized around the stationary operating point and solved in the frequency domain. The resulting small-signal analysis yields the complex displacement, pressure, and velocity fields as a function of frequency.\\
Thermoviscous acoustics were used to describe the air motion inside the beam--stator cavities and outlet structures. In contrast to pressure acoustics, the thermoviscous formulation explicitly resolves viscous and thermal boundary-layer effects and is therefore suitable for the small characteristic dimensions of the investigated microstructures. Fluid--structure interaction was applied at all moving solid--fluid boundaries, allowing the oscillating microbeams to generate acoustic volume flow within the surrounding air domain.\\
Small exterior air domains were additionally included above and below the outlet openings. These regions capture thermoviscous dissipation occurring at the acoustic interfaces and provide the coupling boundary to the exterior radiation model described in the following subsection.\\
The dimensions of the electrostatic gaps place the flow regime close to the transition between continuum and slip-flow conditions. Nevertheless, no-slip boundary conditions were assumed throughout the fluid domain. Consequently, slip-flow effects may influence the fluid dynamics inside the beam--stator cavities. Since no-slip boundary conditions are assumed, the predicted damping may be overestimated in the narrowest regions of the device.\\
The thermoviscous FEM model provides the internal pressure distribution, velocity field, acoustic volume flow, actuator displacement, and electromechanical response of the transducer. These quantities form the basis for the coupled FEM--BEM simulations used to predict the radiated acoustic field.

\subsection{FEM--BEM coupling}
While the thermoviscous FEM model resolves the internal electromechanical and fluidic behavior of a single actuator cell, prediction of the radiated sound field requires consideration of the surrounding acoustic domain. To this end, the thermoviscous FEM model was coupled to an exterior acoustic model based on the boundary element method (BEM) using the COMSOL Acoustics Module.\\
The BEM geometry was derived from the CAD representation of the investigated assembly and included the MEMS chip, printed circuit board, socket, and horn when applicable. Geometric features whose dimensions were considered negligible compared to the acoustic wavelength were omitted to reduce computational complexity. The resulting numerical models are shown in Fig.~\ref{fig:first_principles}d. Owing to geometric symmetry, only half of the assembly was considered during the calculations.\\
The coupling between the thermoviscous FEM domain and the exterior BEM domain was based on the acoustic volume flow exiting the transducer. The normal velocity imposed on the BEM radiation surface was determined from the simulated actuator-cell velocity field and scaled according to the number of active unit cells:
\begin{equation}
v_n^{\mathrm{BEM}} = n_{\mathrm{act}} \frac{\int_{\Gamma_{\mathrm{FEM}}} v_n^{\mathrm{FEM}} \,\mathrm{d}\Gamma_{\mathrm{FEM}}} {\int_{\Gamma_{\mathrm{BEM}}} \mathrm{d}\Gamma_{\mathrm{BEM}}},
\end{equation}
with
\begin{equation}
v_n=\mathbf{v}\cdot\mathbf{n}_{\Gamma},
\end{equation}
where $n_{\mathrm{act}}$ denotes the number of active actuator cells and $\mathbf{n}_{\Gamma}$ the outward surface normal.\\
The pressure obtained from the BEM solution was subsequently transferred back to the thermoviscous FEM model using the spatially averaged pressure over the BEM coupling surface:
\begin{equation}
p^{\mathrm{FEM}} = \frac{\int_{\Gamma_{\mathrm{BEM}}} p^{\mathrm{BEM}} \,\mathrm{d}\Gamma_{\mathrm{BEM}}} {\int_{\Gamma_{\mathrm{BEM}}} \mathrm{d}\Gamma_{\mathrm{BEM}}}.
\end{equation}
This procedure establishes a bidirectional coupling between the internal thermoviscous dynamics and the exterior acoustic radiation field.\\
The approach is motivated by Huygens' principle, where the collective radiation of the closely spaced outlet structures gives rise to an effective continuous wavefront. Nevertheless, this representation remains an approximation because the local radiation impedance experienced by individual actuator cells may vary across the chip area. For example, outlet openings located near the chip edge may experience different acoustic loading than those positioned close to the chip center.\\
The coupled FEM--BEM framework provides the radiated acoustic pressure field, directivity characteristics, and sound-power estimates used throughout the present work.

\subsection{Measurement setup}
The acoustic measurements were performed using the experimental setup shown in Fig.~\ref{fig:first_principles}c. The L-CMUT was mounted on a custom printed circuit board and positioned at the center of a spherical scanning system. Acoustic pressures were measured using a calibrated microphone mounted on a computer-controlled positioning system, enabling automated acquisition of pressure data on a spherical measurement surface surrounding the transducer.\\
Throughout this work, the transducer was operated using a DC bias voltage of \unit[30]{V} and a harmonic excitation voltage with an amplitude of \unit[10]{V}. Identical operating conditions were employed for the numerical simulations in order to facilitate direct comparison between predicted and measured acoustic responses.\\
The excitation signal was generated using an NTi Flexus audio analyzer and amplified by a Krohn-Hite 7602M amplifier. Frequency sweeps were performed between \unit[10]{kHz} and \unit[60]{kHz}. For each measurement position, the microphone signal and a reference signal derived from the excitation were recorded simultaneously. The amplitude and phase of the acoustic response were subsequently determined relative to the excitation signal and used for the spherical near-field acoustic holography reconstruction.\\
Measurements were conducted for both the unloaded transducer and the horn-loaded configuration. The acquired pressure distributions served as the input for the spherical near-field acoustic holography procedure described in the following subsection.

\subsection{Spherical near-field acoustic holography} The radiated acoustic field was reconstructed using spherical near-field acoustic holography (NAH) based on a spherical harmonic representation of the measured sound field \cite{Williams.1999}. Radiation from sources of finite spatial extent enclosed by a spherical measurement surface can be represented as a superposition of outgoing spherical wave components. Once the corresponding spherical harmonic coefficients are known, the acoustic pressure field can be evaluated at arbitrary positions outside the source region.\\
For a given angular frequency $\omega$, the acoustic pressure field is represented as \begin{equation} p \left( r,\theta,\varphi,\omega \right) = \sum_{n=0}^{\infty} \sum_{m=-n}^{n} C_{mn}(\omega) h_n^{(1)}(kr) Y_n^m(\theta,\varphi), \end{equation} where $C_{mn}$ are the spherical harmonic coefficients, $h_n^{(1)}(kr)$ denotes the spherical Hankel function of the first kind, $k$ is the acoustic wavenumber, and $Y_n^m(\theta,\varphi)$ are the spherical harmonics \cite{Williams.1999}
\begin{equation}
Y_n^m(\theta,\varphi) = \sqrt{\frac{2n+1}{4\pi} \frac{(n-m)!}{(n+m)!}} P_n^m(\cos\theta) e^{im\varphi},
\end{equation}
where $\theta\in[0,\pi]$ denotes the polar angle, $\varphi\in[0,2\pi]$ the azimuthal angle, $n\geq0$ the harmonic order, and $|m|\leq n$ the harmonic degree. Furthermore, $P_n^m$ are the associated Legendre polynomials
\begin{equation}
P_n^m(x) = (-1)^m (1-x^2)^{m/2} \frac{\mathrm{d}^m}{\mathrm{d}x^m} P_n(x),
\end{equation}
with
\begin{equation}
P_n(x) = \frac{1}{2^n n!} \frac{\mathrm{d}^n}{\mathrm{d}x^n} (x^2-1)^n
\end{equation}
being the Legendre polynomials.\\
When the acoustic pressure is known on a spherical measurement surface of radius $r=a$, the spherical harmonic coefficients can be determined using the orthogonality of the spherical harmonics \cite{Williams.1999}
\begin{equation}
C_{mn}(\omega) = \frac{1} {h_n^{(1)}(ka)} \int_{\Omega} p(a,\theta,\varphi,\omega) Y_n^m(\theta,\varphi)^* \,\mathrm{d}\Omega,
\end{equation}
where
\begin{equation}
\mathrm{d}\Omega = \sin\theta \,\mathrm{d}\theta \,\mathrm{d}\varphi
\end{equation}
denotes the differential solid angle and $(\cdot)^*$ the complex conjugate. In practice, the integral was evaluated using the discrete pressure measurements obtained on the spherical scanning surface.\\
The spherical harmonic coefficients were subsequently employed for reconstruction of the radiated sound field and determination of the radiated acoustic power. Following Williams \cite{Williams.1999}, the radiated power associated with an individual spherical harmonic component is given by
\begin{equation}
\Pi_{mn} = \frac{1} {2\rho_0 c k^2} \left| C_{mn} \right|^2,
\end{equation}
where $\rho_0$ is the ambient air density and $c$ is the speed of sound. The total radiated acoustic power is obtained as
\begin{equation}
\Pi = \sum_{n=0}^{\infty} \sum_{m=-n}^{n} \Pi_{mn}.
\label{eq:Power_sum}
\end{equation}
The symbol $\Pi$ is used for acoustic power in accordance with Ref.~\cite{Williams.1999} to avoid confusion with the Legendre polynomials $P_n$.\\
For practical implementation, the spherical harmonic expansion was truncated at a finite maximum order $N$. The influence of truncation order on reconstruction accuracy was evaluated through convergence of the radiated acoustic power. The selected truncation order was subsequently used for all field reconstructions, radiated power calculations, and spherical harmonic analyses presented in the Results section

\subsection{Determination of radiated acoustic power}
Radiated acoustic power was evaluated using the spherical harmonic coefficients obtained either from the holographic reconstruction or directly from the FEM--BEM simulations. For the experimental investigations, the coefficients were determined from the measured pressure distribution on the spherical scanning surface as described in the previous subsection. The resulting coefficients were subsequently used to calculate the radiated sound power according to Eq.~\eqref{eq:Power_sum}.\\
For the numerical investigations, the same procedure was applied to the pressure field predicted by the coupled FEM--BEM model. This approach ensured that radiated power was determined consistently for both measurement and simulation, thereby enabling direct comparison between the two datasets.\\
Since practical implementations require truncation of the spherical harmonic expansion at a finite order $N$, the influence of truncation order on the calculated radiated power was evaluated systematically. Radiated power values were computed for increasing values of $N$ and compared with the corresponding FEM--BEM reference solutions. The maximum harmonic order was subsequently selected such that further increases produced only negligible changes in the resulting power estimates.\
The same coefficient sets were additionally used for reconstruction of the directivity patterns and for the spherical harmonic analyses presented in the Results section. Consequently, the radiated sound power, directivity characteristics, and spherical harmonic content were obtained from a common mathematical representation of the measured and simulated acoustic fields.

\section*{Author contributions}
A.M. conceived and designed the study, developed the multiphysics models, performed the numerical simulations, analyzed the data, prepared the figures, and wrote the manuscript. A.A. performed the acoustic measurements and contributed to measurement data acquisition and validation. T.M. contributed to the horn concept used in this work and reviewed the manuscript. J.M. contributed to the original L-CMUT concept and reviewed the manuscript. All authors discussed the results, revised the manuscript, and approved the final version.

\section*{Competing interests}
The authors declare no competing interests.

\subsection*{Data Availability} The datasets generated and analyzed during the current study are not publicly available due to institutional and intellectual property restrictions but may be available from the corresponding author upon reasonable request and with permission from Fraunhofer IPMS.


\begin{thebibliography}{10}
\expandafter\ifx\csname url\endcsname\relax
  \def\url#1{\texttt{#1}}\fi
\expandafter\ifx\csname urlprefix\endcsname\relax\def\urlprefix{URL }\fi
\providecommand{\bibinfo}[2]{#2}
\providecommand{\eprint}[2][]{\url{#2}}

\bibitem{Gerardo.2018}
\bibinfo{author}{Gerardo, C.~D.}, \bibinfo{author}{Cretu, E.} \& \bibinfo{author}{Rohling, R.}
\newblock \bibinfo{title}{{Fabrication and testing of polymer-based capacitive micromachined ultrasound transducers for medical imaging}}.
\newblock \emph{\bibinfo{journal}{{Microsystems {\&} Nanoengineering}}} \textbf{\bibinfo{volume}{4}}, \bibinfo{pages}{19} (\bibinfo{year}{2018}).
\newblock \urlprefix\url{https://www.nature.com/articles/s41378-018-0022-5}.

\bibitem{KhuriYakub.2011}
\bibinfo{author}{Khuri-Yakub, B.~T.} \& \bibinfo{author}{Oralkan, O.}
\newblock \bibinfo{title}{{Capacitive micromachined ultrasonic transducers for medical imaging and therapy}}.
\newblock \emph{\bibinfo{journal}{{Journal of Micromechanics and Microengineering}}} \textbf{\bibinfo{volume}{21}}, \bibinfo{pages}{54004--54014} (\bibinfo{year}{2011}).
\newblock \urlprefix\url{https://iopscience.iop.org/article/10.1088/0960-1317/21/5/054004/meta}.

\bibitem{Brenner.2019}
\bibinfo{author}{Brenner, K.} \emph{et~al.}
\newblock \bibinfo{title}{{Advances in Capacitive Micromachined Ultrasonic Transducers}}.
\newblock \emph{\bibinfo{journal}{{Micromachines}}} \textbf{\bibinfo{volume}{10}} (\bibinfo{year}{2019}).

\bibitem{Wang.2019b}
\bibinfo{author}{Wang, Z.} \emph{et~al.}
\newblock \bibinfo{title}{{Hand Gesture Recognition Based on Active Ultrasonic Sensing of Smartphone: A Survey}}.
\newblock \emph{\bibinfo{journal}{{IEEE Access}}} \textbf{\bibinfo{volume}{7}}, \bibinfo{pages}{111897--111922} (\bibinfo{year}{2019}).

\bibitem{Gijsenbergh.2019}
\bibinfo{author}{Gijsenbergh, P.} \emph{et~al.}
\newblock \bibinfo{title}{{Characterization of polymer-based piezoelectric micromachined ultrasound transducers for short-range gesture recognition applications}}.
\newblock \emph{\bibinfo{journal}{{Journal of Micromechanics and Microengineering}}} \textbf{\bibinfo{volume}{29}}, \bibinfo{pages}{074001} (\bibinfo{year}{2019}).
\newblock \urlprefix\url{https://iopscience.iop.org/article/10.1088/1361-6439/ab1f41}.

\bibitem{Yaralioglu.2003}
\bibinfo{author}{Yaralioglu, G.~G.}, \bibinfo{author}{Ergun, A.~S.}, \bibinfo{author}{Huang, Y.} \& \bibinfo{author}{Khuri-Yakub, B.~T.}
\newblock \bibinfo{title}{{Capacitive micromachined ultrasonic transducers for robotic sensing applications}}.
\newblock In \emph{\bibinfo{booktitle}{{Proceedings / 2003 IEEE/RSJ International Conference on Intelligent Robots and Systems (IROS 2003)}}}, \bibinfo{pages}{2347--2352} (\bibinfo{publisher}{{IEEE Operations Center}}, \bibinfo{address}{Piscataway, NJ}, \bibinfo{year}{2003}).

\bibitem{Yu.2016}
\bibinfo{author}{Yu, J.}, \bibinfo{author}{Li, S.~E.}, \bibinfo{author}{Liu, C.} \& \bibinfo{author}{Cheng, B.}
\newblock \bibinfo{title}{{Dynamical tracking of surrounding objects for road vehicles using linearly-arrayed ultrasonic sensors}}.
\newblock In \emph{\bibinfo{booktitle}{{2016 IEEE Intelligent Vehicles Symposium (IV)}}}, \bibinfo{pages}{72--77} (\bibinfo{publisher}{IEEE}, \bibinfo{address}{Piscataway, NJ}, \bibinfo{year}{2016}).

\bibitem{Przybyla.2015}
\bibinfo{author}{Przybyla, R.~J.} \emph{et~al.}
\newblock \bibinfo{title}{{3D Ultrasonic Rangefinder on a Chip}}.
\newblock \emph{\bibinfo{journal}{{IEEE Journal of Solid-State Circuits}}} \textbf{\bibinfo{volume}{50}}, \bibinfo{pages}{320--334} (\bibinfo{year}{2015}).

\bibitem{Unger.2019}
\bibinfo{author}{Unger, A.}
\newblock \emph{\bibinfo{title}{{Air Coupled Ultrasonic Transducers for Industrial Applications}}}.
\newblock Ph.D. thesis, \bibinfo{address}{Darmstadt}.
\newblock \urlprefix\url{https://tuprints.ulb.tu-darmstadt.de/8974/}.

\bibitem{Park.2013}
\bibinfo{author}{Park, K.~K.} \& \bibinfo{author}{Khuri-Yakub, B.~T.}
\newblock \bibinfo{title}{{3-D airborne ultrasound synthetic aperture imaging based on capacitive micromachined ultrasonic transducers}}.
\newblock \emph{\bibinfo{journal}{{Ultrasonics}}} \textbf{\bibinfo{volume}{53}}, \bibinfo{pages}{1355--1362} (\bibinfo{year}{2013}).
\newblock \urlprefix\url{https://www.sciencedirect.com/science/article/pii/S0041624X1300098X}.

\bibitem{claudon_advances_2002}
\bibinfo{author}{Claudon, M.}, \bibinfo{author}{Tranquart, F.}, \bibinfo{author}{Evans, D.~H.}, \bibinfo{author}{Lefèvre, F.} \& \bibinfo{author}{Correas, J.}
\newblock \emph{\bibinfo{title}{Advances in ultrasound}}, vol.~\bibinfo{volume}{12} (\bibinfo{year}{2002}).
\newblock \urlprefix\url{https://doi.org/10.1007/s00330-001-1185-1}.

\bibitem{bond_ultrasonics_2024}
\bibinfo{author}{Bond, D.~E., Leonard~J.}
\newblock \emph{\bibinfo{title}{Ultrasonics: {Fundamentals}, {Technologies}, and {Applications}}} (\bibinfo{publisher}{CRC Press}, \bibinfo{address}{Boca Raton}, \bibinfo{year}{2024}), \bibinfo{edition}{4} edn.

\bibitem{lacefield_ultrasound_2015}
\bibinfo{editor}{Lacefield, A.~F., James~C.} (ed.) \emph{\bibinfo{title}{Ultrasound {Imaging} and {Therapy}}} (\bibinfo{publisher}{CRC Press}, \bibinfo{address}{Boca Raton}, \bibinfo{year}{2015}).

\bibitem{yamamoto_holographic_1984}
\bibinfo{author}{Yamamoto, T.} \& \bibinfo{author}{Aoki, Y.}
\newblock \bibinfo{title}{Holographic {B}-{Scope} {Imaging} {Using} {Wide}-{Band} {Chirped} {Ultrasound}}.
\newblock \emph{\bibinfo{journal}{IEEE Transactions on Sonics and Ultrasonics}} \textbf{\bibinfo{volume}{31}}, \bibinfo{pages}{362--366} (\bibinfo{year}{1984}).
\newblock \urlprefix\url{http://ieeexplore.ieee.org/document/1539592/}.

\bibitem{bente_broadband_2023}
\bibinfo{author}{Bente, K.}, \bibinfo{author}{Rus, J.}, \bibinfo{author}{Mooshofer, H.}, \bibinfo{author}{Gaal, M.} \& \bibinfo{author}{Grosse, C.~U.}
\newblock \bibinfo{title}{Broadband {Air}-{Coupled} {Ultrasound} {Emitter} and {Receiver} {Enable} {Simultaneous} {Measurement} of {Thickness} and {Speed} of {Sound} in {Solids}}.
\newblock \emph{\bibinfo{journal}{Sensors}} \textbf{\bibinfo{volume}{23}}, \bibinfo{pages}{1379} (\bibinfo{year}{2023}).
\newblock \urlprefix\url{https://www.mdpi.com/1424-8220/23/3/1379}.
\newblock \bibinfo{note}{Number: 3 Publisher: Multidisciplinary Digital Publishing Institute}.

\bibitem{otsuka_non-contact_2022}
\bibinfo{author}{Otsuka, H.} \& \bibinfo{author}{Okubo, K.}
\newblock \bibinfo{title}{Non-contact identification of moisture content of fabric based on analysis of broadband acoustic signals using multiple-frequency air ultrasonic transducer system}.
\newblock \emph{\bibinfo{journal}{Japanese Journal of Applied Physics}} \textbf{\bibinfo{volume}{61}}, \bibinfo{pages}{SG8003} (\bibinfo{year}{2022}).
\newblock \urlprefix\url{https://dx.doi.org/10.35848/1347-4065/ac51c3}.
\newblock \bibinfo{note}{Publisher: IOP Publishing}.

\bibitem{Muralt.2005}
\bibinfo{author}{Muralt, P.} \emph{et~al.}
\newblock \bibinfo{title}{{Piezoelectric micromachined ultrasonic transducers based on PZT thin films}}.
\newblock \emph{\bibinfo{journal}{{IEEE transactions on ultrasonics, ferroelectrics, and frequency control}}} \textbf{\bibinfo{volume}{52}}, \bibinfo{pages}{2276--2288} (\bibinfo{year}{2005}).

\bibitem{Akasheh.2005}
\bibinfo{author}{Akasheh, F.}, \bibinfo{author}{Fraser, J.~D.}, \bibinfo{author}{Bose, S.} \& \bibinfo{author}{Bandyopadhyay, A.}
\newblock \bibinfo{title}{{Piezoelectric micromachined ultrasonic transducers: modeling the influence of structural parameters on device performance}}.
\newblock \emph{\bibinfo{journal}{{IEEE transactions on ultrasonics, ferroelectrics, and frequency control}}} \textbf{\bibinfo{volume}{52}}, \bibinfo{pages}{455--468} (\bibinfo{year}{2005}).

\bibitem{Przybyla2012}
\bibinfo{author}{Przybyla, R.~J.} \emph{et~al.}
\newblock \bibinfo{title}{In-air ultrasonic rangefinding and angle estimation using an array of {AlN} micromachined transducers}.
\newblock In \emph{\bibinfo{booktitle}{2012 Solid-State, Actuators, and Microsystems Workshop}} (\bibinfo{year}{2012}).

\bibitem{Salim.2012}
\bibinfo{author}{Salim, M.~S.}, \bibinfo{author}{{Abd Malek}, M.~F.}, \bibinfo{author}{Heng, R.}, \bibinfo{author}{Juni, K.~M.} \& \bibinfo{author}{Sabri, N.}
\newblock \bibinfo{title}{{Capacitive Micromachined Ultrasonic Transducers: Technology and Application}}.
\newblock \emph{\bibinfo{journal}{{Journal of Medical Ultrasound}}} \textbf{\bibinfo{volume}{20}}, \bibinfo{pages}{8--31} (\bibinfo{year}{2012}).

\bibitem{Kupnik2011}
\bibinfo{author}{Kupnik, M.}, \bibinfo{author}{Ho, M.-C.}, \bibinfo{author}{Vaithilingam, S.} \& \bibinfo{author}{Khuri-Yakub, B.~T.}
\newblock \bibinfo{title}{{CMUTs} for air coupled ultrasound with improved bandwidth}.
\newblock In \emph{\bibinfo{booktitle}{2011 IEEE International Ultrasonics Symposium}}, \bibinfo{pages}{592--595} (\bibinfo{year}{2011}).

\bibitem{ZhangH2020}
\bibinfo{author}{Zhang, H.} \emph{et~al.}
\newblock \bibinfo{title}{Fabrication and characterization of a wideband low-frequency {CMUT} array for air-coupled imaging}.
\newblock \emph{\bibinfo{journal}{IEEE Sensors Journal}} \textbf{\bibinfo{volume}{20}}, \bibinfo{pages}{14090--14100} (\bibinfo{year}{2020}).

\bibitem{Wygant50kHz}
\bibinfo{author}{Wygant, I.~O.} \emph{et~al.}
\newblock \bibinfo{title}{50 {kHz} capacitive micromachined ultrasonic transducers for generation of highly directional sound with parametric arrays}.
\newblock \emph{\bibinfo{journal}{IEEE Transactions on Ultrasonics, Ferroelectrics, and Frequency Control}} \textbf{\bibinfo{volume}{56}}, \bibinfo{pages}{193--203} (\bibinfo{year}{2009}).

\bibitem{kang_wideband_2019}
\bibinfo{author}{Kang, L.}, \bibinfo{author}{Feeney, A.} \& \bibinfo{author}{Dixon, S.}
\newblock \bibinfo{title}{Wideband electromagnetic dynamic acoustic transducers ({WEMDATs}) for air-coupled ultrasonic applications}.
\newblock \emph{\bibinfo{journal}{Applied Physics Letters}} \textbf{\bibinfo{volume}{114}}, \bibinfo{pages}{053505} (\bibinfo{year}{2019}).
\newblock \urlprefix\url{https://doi.org/10.1063/1.5086383}.

\bibitem{ealo_broadband_2008}
\bibinfo{author}{Ealo, J.}, \bibinfo{author}{Seco, F.} \& \bibinfo{author}{Jimenez, A.}
\newblock \bibinfo{title}{Broadband {EMFi}-based transducers for ultrasonic air applications}.
\newblock \emph{\bibinfo{journal}{IEEE Transactions on Ultrasonics, Ferroelectrics and Frequency Control}} \textbf{\bibinfo{volume}{55}}, \bibinfo{pages}{919--929} (\bibinfo{year}{2008}).
\newblock \urlprefix\url{http://ieeexplore.ieee.org/document/4494787/}.

\bibitem{sun_investigation_2019}
\bibinfo{author}{Sun, C.} \emph{et~al.}
\newblock \bibinfo{title}{Investigation of {Broadband} {Characteristics} of {Multi}-{Frequency} {Piezoelectric} {Micromachined} {Ultrasonic} {Transducer} ({MF}-{pMUT})}.
\newblock \emph{\bibinfo{journal}{IEEE Sensors Journal}} \textbf{\bibinfo{volume}{19}}, \bibinfo{pages}{860--867} (\bibinfo{year}{2019}).
\newblock \urlprefix\url{https://ieeexplore.ieee.org/document/8515238/}.

\bibitem{wang_multi-frequency_2023}
\bibinfo{author}{Wang, Z.}, \bibinfo{author}{Duan, H.} \& \bibinfo{author}{Zheng, Y.}
\newblock \bibinfo{title}{Multi-frequency broadband piezoelectric micromachined ultrasonic transducer utilizing helmholtz resonance}.
\newblock \emph{\bibinfo{journal}{International Journal of Mechanical Sciences}} \textbf{\bibinfo{volume}{249}}, \bibinfo{pages}{108267} (\bibinfo{year}{2023}).
\newblock \urlprefix\url{https://www.sciencedirect.com/science/article/pii/S0020740323001698}.

\bibitem{Apte2013}
\bibinfo{author}{Apte, N.}, \bibinfo{author}{Park, K.~K.} \& \bibinfo{author}{Khuri-Yakub, B.~T.}
\newblock \bibinfo{title}{Experimental evaluation of {CMUTs} with vented cavities under varying pressure}.
\newblock In \emph{\bibinfo{booktitle}{2013 IEEE International Ultrasonics Symposium (IUS)}}, \bibinfo{pages}{1724--1727} (\bibinfo{year}{2013}).

\bibitem{MaCMUT}
\bibinfo{author}{Ma, B.}, \bibinfo{author}{Firouzi, K.}, \bibinfo{author}{Brenner, K.} \& \bibinfo{author}{Khuri-Yakub, B.~T.}
\newblock \bibinfo{title}{Wide bandwidth and low driving voltage vented cmuts for airborne applications}.
\newblock \emph{\bibinfo{journal}{IEEE Transactions on Ultrasonics, Ferroelectrics, and Frequency Control}} \textbf{\bibinfo{volume}{66}}, \bibinfo{pages}{1777--1785} (\bibinfo{year}{2019}).

\bibitem{Park2021}
\bibinfo{author}{Park, B.~H.}, \bibinfo{author}{Kim, B.~K.} \& \bibinfo{author}{Park, K.~K.}
\newblock \bibinfo{title}{Fractional bandwidth improvement of ultrasonic airborne transducers using acoustic block}.
\newblock \emph{\bibinfo{journal}{Journal of Mechanical Science and Technology}} \textbf{\bibinfo{volume}{35}} (\bibinfo{year}{2021}).

\bibitem{kuntzman_microfabrication_2015}
\bibinfo{author}{Kuntzman, M.~L.}, \bibinfo{author}{Kim, D.} \& \bibinfo{author}{Hall, N.~A.}
\newblock \bibinfo{title}{Microfabrication and {Experimental} {Evaluation} of a {Rotational} {Capacitive} {Micromachined} {Ultrasonic} {Transducer}}.
\newblock \emph{\bibinfo{journal}{Journal of Microelectromechanical Systems}} \textbf{\bibinfo{volume}{24}}, \bibinfo{pages}{404--413} (\bibinfo{year}{2015}).
\newblock \urlprefix\url{http://ieeexplore.ieee.org/document/6856137/}.

\bibitem{Adelegan_Spiral}
\bibinfo{author}{Adelegan, O.~J.}, \bibinfo{author}{Coutant, Z.~A.}, \bibinfo{author}{Wu, X.}, \bibinfo{author}{Yamaner, F.~Y.} \& \bibinfo{author}{Oralkan, {\"O}.}
\newblock \bibinfo{title}{Design and fabrication of wideband air-coupled capacitive micromachined ultrasonic transducers with varying width annular-ring and spiral cell structures}.
\newblock \emph{\bibinfo{journal}{IEEE Transactions on Ultrasonics, Ferroelectrics, and Frequency Control}} \bibinfo{pages}{1--1} (\bibinfo{year}{2021}).

\bibitem{zhang_fabrication_2020}
\bibinfo{author}{Zhang, H.} \emph{et~al.}
\newblock \bibinfo{title}{Fabrication and {Characterization} of a {Wideband} {Low}-{Frequency} {CMUT} {Array} for {Air}-{Coupled} {Imaging}}.
\newblock \emph{\bibinfo{journal}{IEEE Sensors Journal}} \textbf{\bibinfo{volume}{20}}, \bibinfo{pages}{14090--14100} (\bibinfo{year}{2020}).
\newblock \urlprefix\url{https://ieeexplore.ieee.org/document/9133447/}.

\bibitem{wang_hybrid_2021}
\bibinfo{author}{Wang, H.} \emph{et~al.}
\newblock \bibinfo{title}{Hybrid {Cell} {Structure} for {Wideband} {CMUT}: {Design} {Method} and {Characteristic} {Analysis}}.
\newblock \emph{\bibinfo{journal}{Micromachines}} \textbf{\bibinfo{volume}{12}}, \bibinfo{pages}{1180} (\bibinfo{year}{2021}).
\newblock \urlprefix\url{https://www.mdpi.com/2072-666X/12/10/1180}.
\newblock \bibinfo{note}{Number: 10 Publisher: Multidisciplinary Digital Publishing Institute}.

\bibitem{lu_theoretical_2022}
\bibinfo{author}{Lu, W.} \emph{et~al.}
\newblock \bibinfo{title}{Theoretical and {Simulation} {Studies} on {Designing} a {Phase}-{Reversal}-{Based} {Broadband} {CMUT} {With} {Flat} {Passband} and {Improved} {Noise} {Rejections} for {SHM}}.
\newblock \emph{\bibinfo{journal}{IEEE Sensors Journal}} \textbf{\bibinfo{volume}{22}}, \bibinfo{pages}{21527--21539} (\bibinfo{year}{2022}).
\newblock \urlprefix\url{https://ieeexplore.ieee.org/document/9911184/}.

\bibitem{Monsalve.2022}
\bibinfo{author}{Monsalve, J.~M.} \emph{et~al.}
\newblock \bibinfo{title}{Proof of concept of an air-coupled electrostatic ultrasonic transducer based on lateral motion}  (\bibinfo{year}{2022}).

\bibitem{Monsalve_2025_LRF}
\bibinfo{author}{Monsalve, J.~M.} \& \bibinfo{author}{Schenk, H. A.~G.}
\newblock \bibinfo{title}{A visco-thermal model for a novel type of electrostatic micromachined in-plane ultrasound transducer}.
\newblock \emph{\bibinfo{journal}{Acta Acustica}} \textbf{\bibinfo{volume}{10}} (\bibinfo{year}{2026}).

\bibitem{Meisel2022}
\bibinfo{author}{Meisel, T.} \emph{et~al.}
\newblock \bibinfo{title}{Directivity optimization of {MEMS} ultrasonic transducers by implementing acoustic horns}.
\newblock In \emph{\bibinfo{booktitle}{Proceedings of the 24th International Congress on Acoustics}}, \bibinfo{number}{ABS-0761} (\bibinfo{year}{2022}).

\bibitem{Melnikov.2021}
\bibinfo{author}{Melnikov, A.} \emph{et~al.}
\newblock \bibinfo{title}{{Coulomb-actuated microbeams revisited: experimental and numerical modal decomposition of the saddle-node bifurcation}}.
\newblock \emph{\bibinfo{journal}{{Microsystems {\&} Nanoengineering}}} \textbf{\bibinfo{volume}{7}} (\bibinfo{year}{2021}).

\bibitem{Schenk.2022}
\bibinfo{author}{Schenk, H. A.~G.} \emph{et~al.}
\newblock \bibinfo{title}{{Coulomb actuated microbeams: A Chebyshev-Edgeworth approach to highly efficient lumped parameter models}}.
\newblock \urlprefix\url{https://arxiv.org/pdf/2201.01627}.

\bibitem{Williams.1999}
\bibinfo{author}{Williams, E.~G.}
\newblock \emph{\bibinfo{title}{{Fourier acoustics: Sound radiation and nearfield acoustical holography}}} (\bibinfo{publisher}{{Academic Press}}, \bibinfo{address}{San Diego, Calif.}, \bibinfo{year}{1999}).

\end{thebibliography}
\end{document}